\documentclass[11pt]{article}
\usepackage{amsmath,amssymb,graphicx,float,epsf}
\def\RR{\hbox{I\kern-.2em\hbox{R}}}
\newcommand{\qed}{\hbox to 0pt{}\hfill$\rlap{$\sqcap$}\sqcup$ \vspace{3mm}}
\numberwithin{equation}{section}
\restylefloat{figure}
\usepackage{graphicx} 
\usepackage{algorithm}
\usepackage{algorithmic}
\usepackage[table]{xcolor}
\usepackage{longtable}
\usepackage{ragged2e}
\usepackage{amsmath}
\usepackage{amsthm}
\usepackage{amsfonts}
\usepackage{ mathrsfs }
\usepackage[caption = false]{subfig}
\usepackage{wrapfig}
\usepackage{enumerate}
\usepackage{enumitem}
\usepackage{array}
\usepackage{subfig}
\usepackage[margin=1cm]{caption}
\usepackage[margin=1in]{geometry}
\usepackage{authblk}

\usepackage[utf8]{inputenc}

\usepackage{graphicx}
\usepackage{hyperref}
\hypersetup{
    colorlinks=true,
    linkcolor=blue,
    filecolor=magenta,
    urlcolor=cyan,
    citecolor=blue,
}
\date{}
\graphicspath{ {Images/} }
\begin{document}

    \title{Downscaling Epidemiological Time Series Data for Improving Forecasting Accuracy: An Algorithmic Approach}

    \author[1,2]{\small Mahadee Al Mobin\thanks{Email: mahadeealmobin@gmail.com}}
    \author[1]{\small Md. Kamrujjaman\thanks{Corresponding author: Md Kamrujjaman, E-mail: kamrujjaman@du.ac.bd}}

    \affil[1]{\footnotesize Department of Mathematics, University of Dhaka, Dhaka 1000, Bangladesh}
    \affil[2]{\footnotesize Bangladesh Institute of Governance and Management, Dhaka, Bangladesh}
    

    \maketitle

    \vspace{-1.0cm}
    \noindent\rule{6.35in}{0.02in}\\
    {\bf Abstract.}
     Data scarcity and discontinuity are common occurrences in the healthcare and epidemiological dataset and often need help in forming an educative decision and forecasting the upcoming scenario. Often, these data are stored as monthly/yearly aggregate where the prevalent forecasting tools like Autoregressive Integrated Moving Average (ARIMA), Seasonal Autoregressive Integrated Moving Average (SARIMA), and TBATS often fail to provide satisfactory results. Artificial data synthesis methods have been proven to be a powerful tool for tackling these challenges. The paper aims to propose a downscaling data algorithm based on the underlying distribution. Our findings show that the synthesized data is in agreement with the original data in terms of trend, seasonality, and residuals, and the synthesized data provides a stable foothold for the forecasting tools to generate a much more accurate forecast of the situation.\\

    \noindent{\it \footnotesize Keywords}: {\small Epidemiology, Downscaling Algorithm, Temporal Downscaling, ARIMA, Fourier-ARIMA, Dengue.}\\
    \noindent

    	\section{Introduction}\label{introduction_1}
    Any process that involves deriving high-resolution data from low-resolution variables is referred to as downscaling. This method relies on dynamical or statistical approaches and is extensively utilized in the field of meteorology, climatology, and remote sensing \cite{peng2017review,ribalaygua2013description}. Significant exploration of the downscaling methods has been done in the field of geology and climatology to enhance the out of existing models like the General Circulation Model (GCM) \cite{kim2022optimizing,bae2015climate,lee2019impact,kim2020intensified,gangopadhyay2005statistical,fowler2007linking}, Regional Climate Model (RCM) \cite{lee2014nonparametric}, Integrated Grid Modeling System (IGMS) \cite{buster2021physical}, System Advisor Model (SAM) \cite{buster2021physical} and to make it usable for the forecast of geographically significant region and time. Several methods has been used to downscale these data such as BCC/RCG-Weather Generators (BCC/RCG-WG) \cite{liu2016comparison,yaoming2004stochastic,liao2013change}, and Statistics Downscaling Model (SDSM)\cite{liu2016comparison,dibike2005hydrologic,khan2006uncertainty,wilby2002sdsm,harpham2005multi,wilby2000streamflow,wetterhall2007seasonality}, Bayesian Model Averaging (BMA) \cite{raftery2003discussion}. Even machine learning methods has been used like Genetic algorithm (GA) \cite{lee2014nonparametric}, K Nearest Neighbourhood Resampling (KNNR) \cite{lee2014nonparametric},  Support Vector Machine (SVM) \cite{liu2016comparison,tripathi2006downscaling,yu2007forecasting,ghosh2008statistical}. Except for the machine learning algorithms, which are methods that are finding their applications in new domains, the rest of the methods are tailored to suit the outputs of the models, as mentioned earlier. \\


    This class of methods has recently been applied in the deaggregation of spatial epidemiological data \cite{matisziw2008downscaling}. But significant work has yet to be done for the temporal downscaling of epidemiological data. Often, the temporal downscaling techniques are classical interpolation techniques that do not do justice to aggregated data. This phenomenon can be well illustrated with an example. Consider the case of monthly Dengue infection data of 2017 from figure \ref{fig:monhtly_distro}, which has been downscaled using linear interpolation by considering the aggregated value as the value of the end date of a month in Figure \ref{fig:prior_distro}. In this case, if we consider the monthly aggregate of the downscaled data, it does not match the original aggregate. This downscaled data, which differs from the original data in such statistical measures, shall result in decisions and knowledge that cannot be far from the truth.\\
    
    This paper aims to propose a novel algorithm named Mahadee-Kamrujjaman Downscaling (MKD) algorithm based on the Bayesian approach that can regenerate downscaled temporal time series of varying time lengths from aggregated data preserving most of the statistical characteristics and the aggregated sum of the original data.\\
    
    The paper is organized as follows. Section 2 describes the data used for the paper and its sources. Section 3 discusses the methodology at length with the proposed MKD algorithm. Section 4 compares the synthesized data with the actual data of two different epidemiological cases (Dengue and COVID-19) in Bangladesh and shows how the MKD algorithm could generate statistically accurate approximate of the actual with very little input in both cases, and discuss the benchmark metric used for evaluating the output. Section 5 shows the improvement of the forecasting accuracy using synthesized data over aggregated data using a statistical forecasting toolbox in the dengue scenario of Bangladesh using the last 12 years of monthly aggregated data, Forecasting model selection procedures, and residuals. Finally, section 6 concludes our paper with an overview of the paper and how our paper has contributed to the existing literature and scopes for improvements and fields of application of the MKD algorithm.

    \section{Data}\label{sec:intro}
    The dengue data from Bangladesh used in this paper are from January 2010 to July 2022 and are collected from DGHS \cite{DGHS2022}, and IEDCR \cite{IEDCR2021}. The COVID-19 data of Bangladesh are from 8 March 2020 to December 2020 and are collected from the WHO data repository \cite{WHO2022}.

    \section{Methodology}\label{sec:methodology}
    The MKD algorithm can be segmented into three sequential parts, as exhibited in Figure \ref{fig:flowdiag_1}. Initially, the algorithm considers a prior distribution to generate synthetic downscaled data. The MKD algorithm considers the aggregated data as the prior distribution of the downscaled data. For example: If we have the monthly epidemiological data of dengue for the year 2017, thus to attain the prior distribution for the downscaled data, we divide the data by 30. The fact is well illustrated in Figures \ref{fig:monhtly_distro} and \ref{fig:prior_distro} attached in the appendix \ref{app:fig}. Figure \ref{fig:monhtly_distro} depicts the monthly distribution of the DENV (Dengue Virus) infection in Bangladesh for the year 2017, and Figure \ref{fig:prior_distro} represents the prior distribution obtained by the method described above.\\
    
    \begin{center}
    	\includegraphics[width=\textwidth]{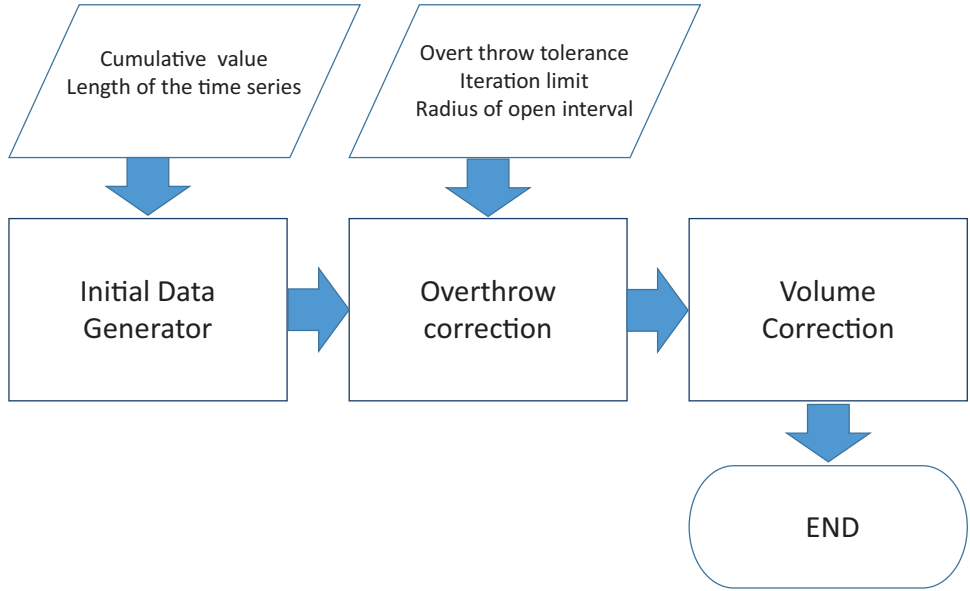}
    	\captionof{figure}{Flow diagram of the MKD algorithm.}
    	\label{fig:flowdiag_1}
    \end{center}

    Based on the prior distribution, initial statistical properties of the synthetic data are obtained except for the standard deviation ($ \sigma $). As $ \sigma $ is scaling independent, hence scaling method used to obtain the prior distribution from the monthly aggregate keeps the $ \sigma $ identical to the monthly aggregate. To overcome this problem, we consider,
    \begin{equation}\label{eq:sigamtransformation}
    	\sigma_0=\displaystyle \frac{\sigma_{prior\,distribution}}{30}
    \end{equation}
    where $ \sigma_0 $ is the standard deviation considered for the distribution to be fitted to generate the downscaled data by the algorithm and $ \sigma_{prior\, distribution} $ is the standard deviation of the obtained prior distribution. Later on, in section \ref{sec:realvssynth}, we will see that the initial assumption of the standard deviation considered in \eqref{eq:sigamtransformation} is a good approximation for the downscaled data.\\

    \subsection{Initial Data Generation}

    The \textit{``Initial Data Generator''} phase feeds on the aggregated data, length of the aggregate interval, and $ \sigma_0 $  to give an initial downscaled data based on a ``Distribution Generator''. Based on the prior distribution, a proper statistical probability distribution (PD) is to be considered to be fitted to generate the data. The ``Distribution Generator'' aims to fit the selected PD to the prior distribution based on the statistical properties obtained for the initial phase. The challenge not only in this scenario but also in every step of the algorithm is to ensure that the synthetic data produced in every step is non-negative integers, as we are dealing with epidemiological data. Thus specific measures have been deployed to tackle these challenges which are:
    \begin{itemize}
    	\item To ensure non negativity consider the transformation: $$ \hat{\mathbf{y}}=\mathbf{y}+min(|\mathbf{y}|) $$

    	\item To ensure that the data points are integer irrespective of the selection of PD, we round off the data to the nearest integer and subtract one from randomly selected data points in each aggregated unit such that the  synthesized data has the same sum as the aggregated unit
    \end{itemize}

    Thus imposing these measures, the ``Distribution Generator'' generates synthetic distribution for each aggregated unit. Thus looping over the entire aggregated timeline generates the initial distribution of the downscaled data with respect to the aggregated data. This initial distribution is a suitable approximation to the actual data but can be improved with further refinement. The synthetic data will result in the exact aggregated data from which it is generated.

    \subsection{Overthrow Correction}\label{subsec:overthrowc}

    This step is often necessary for time series data with an abrupt change in gradient or in case of initial approximation with abnormally large overthrow as the approximations are probabilistic. In case of data with the abrupt change in gradient, the initial approximation is often left with a staircase-like structure as exhibited in the Figure \ref{fig:dengue_2109_overthrow_presmoothhing}. The problem can be corrected using the overthrow correction measure, which is demonstrated in Figure \ref{fig:dengue_2109_overthrow_postsmoothhing}.

    \begin{center}
    	\includegraphics[width=\textwidth]{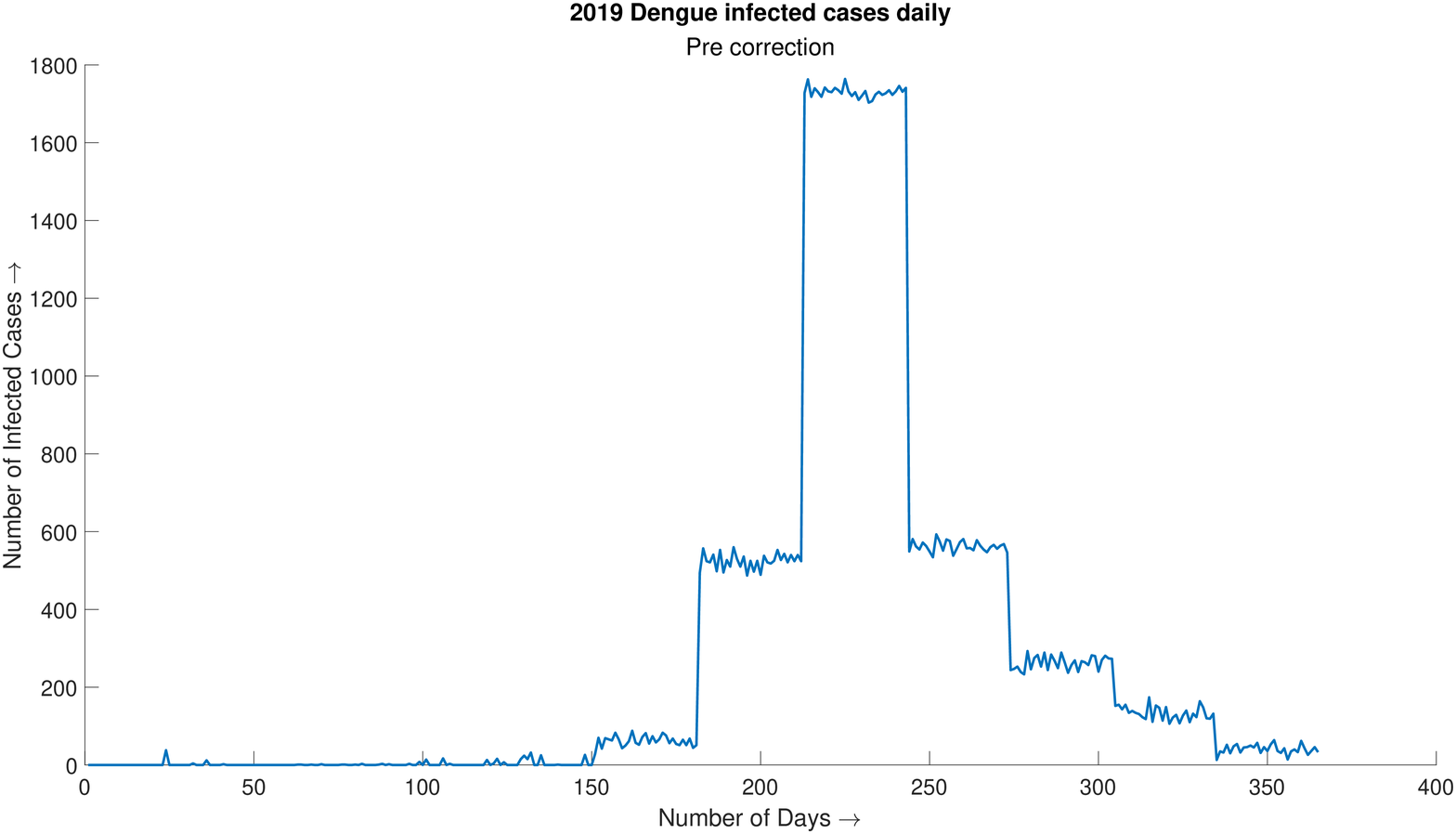}
    	\captionof{figure}{Initial approximation without overthrow correction exhibits a staircase like property due to higher gradient change of the prior distribution.}
    	\label{fig:dengue_2109_overthrow_presmoothhing}
    \end{center}

    The overthrow correction part takes a tolerance, $ \delta $, iteration limit, n, and a radius of an open interval, r. The step initially determine overthrow using tolerance between two neighboring points i.e. if $ y_{i}-y_{i-1}>\delta $ or $ y_{i}-y_{i+1}>\delta  $ then $ y_i $ is an overthrow. After identifying an overthrow, we consider an open interval of radius r around the overthrow point and execute the distribution generator on that open interval. This redistributes the sample within the open interval diminishing the overthrow to some extent. This process is iterated n times over the entire time series to ensure satisfactory results. The strength of the overthrow correction step can be dictated by the two parameters $ \delta $ and n. The strength of the overthrow correction is directly proportional to n and is inversely proportional to $ \delta $. Selecting the correct parameter value can ensure a good approximation of the real-life scenario.
    \newpage
    \begin{center}
    	\includegraphics[width=\textwidth]{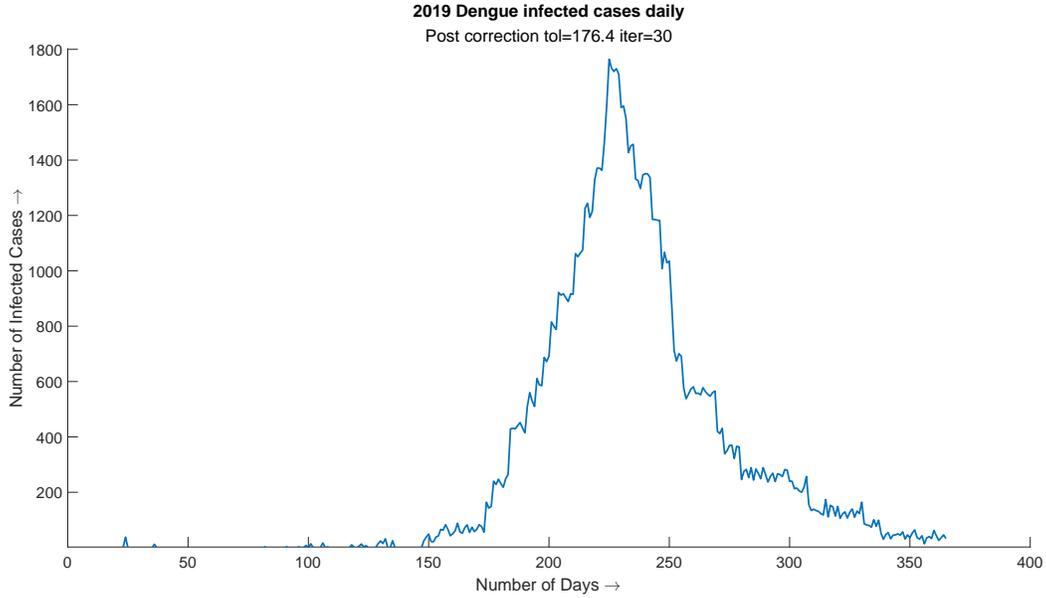}
    	\captionof{figure}{ Initial approximation with overthrow correction exhibits a much proper approximation of the real case scenario preserving its original trend.}
    	\label{fig:dengue_2109_overthrow_postsmoothhing}
    \end{center}


    \subsection{Volume Correction}\label{subsec:volcorr}

    The overthrow correction disrupts the property of the synthesized time series to conserve its aggregated sum equal to the given aggregated distribution due to its local correction property. The scenario best illustrates the table \ref{tab:volumetable}. This problem is addressed in this step. To maintain aggregated sum equal to the original data, we consider each aggregated unit and adjust the sum accordingly, adding/subtracting 1 from randomly chosen indices until the sum equates as required.
    \newpage
    \begin{center}
    	\centering
    	\begin{tabular}{|c|c|>{\centering\arraybackslash}p{2.5cm}|>{\centering\arraybackslash}p{2.5cm}|>{\centering\arraybackslash}p{2.5cm}|}
    		\hline
    		\rowcolor{blue!20}\Centering\bfseries Month & \Centering\bfseries Actual & \Centering\bfseries Initial Distribution & \Centering\bfseries Overthrow Correction &\Centering\bfseries Volume Correction  \\ \hline
    		January & 38 & 38 & 42 & 38  \\ \hline
    		February & 18 & 18 & 13 & 18  \\ \hline
    		March & 17 & 17 & 18 & 17  \\ \hline
    		April & 58 & 58 & 61 & 58  \\ \hline
    		May & 193 & 193 & 300 & 193  \\ \hline
    		June & 1884 & 1884 & 2500 & 1884  \\ \hline
    		July & 16253 & 16253 & 17617 & 16253  \\ \hline
    		August & 53636 & 53636 & 49581 & 53636  \\ \hline
    		September & 16856 & 16856 & 18259 & 16856  \\ \hline
    		October & 8143 & 8143 & 8419 & 8143  \\ \hline
    		November & 4011 & 4011 & 4094 & 4011  \\ \hline
    		December & 1247 & 1247 & 1450 & 1247  \\ \hline
    		\bfseries Total & \textbf{102354} & \textbf{102354} & \textbf{102354} & \textbf{102354}  \\ \hline
    	\end{tabular}
    	\captionof{table}{The table exhibits the comparison of the number of cases each month for executing the MKD algorithm on the Dengue 2019 data of Bangladesh with the actual data. Here we can see the total number of infected individuals in each algorithm step is the same. In the case of the monthly sum, we see some anomaly in the overthrow correction case, which has been fixed in the volume correction step.}
    	\label{tab:volumetable}
    \end{center}
    \subsection{The  Mahadee-Kamrujjaman Downscaling (MKD) Algorithm}
    The algorithms calls for a unique name of it's own. From now on, we shall address it as Mahadee-Kamrujjaman Downscaling (MKD) algorithm. The structural part of the algorithm has been discussed at length to in the first three subsections of the section methodology. The proper pseudo code of the MKD algorithm is as follows:

    \begin{center}
    	\hrule
    	\vspace{0.2in}
    	\captionof{algorithm}{Mahadee-Kamrujjaman Downscaling (MKD) Algorithm}
    	\hrule
    	\vspace{0.2in}

    	\begin{algorithmic}
    		\REQUIRE Aggregated value vector, \textbf{v}\\
    		Overthrow tolerance, $ \delta $\\
    		Iteration limit, n\\
    		Radius of the open interval, r\\
    		Standard deviation, $ \sigma $
    		\ENSURE downscaled time series, $ \bar{v} $
    		\FOR{elem in \textbf{v}}
    		\STATE $ \bar{v} =$ Distribution Generator(elem,$ \sigma $)\ENDFOR
    		\FOR{i from 1 to n}
    		\STATE find a vector of coordinates of overthrow points
    		\FOR{elem in overthrow points}
    		\STATE open interval centering elem of radius, r = Distribution Generator(sum of the elements of open interval,$ \sigma $)
    		\ENDFOR
    		\ENDFOR
    		\FOR {elem in \textbf{v}}
    		\IF{$ v_i \neq$sum of euiquivalent aggregate in $ \bar{v} $}
    		\STATE d=$ v_i $-sum of equivalent aggregate in $ \bar{v} $
    		\WHILE{$ d\neq0 $}
    		\IF{$ d>0 $}
    		\STATE $ \bar{v}_{randomly\,picked\,index}=\bar{v}_{randomly\,picked\,index}+1 $
    		\STATE $ d-=1 $
    		\ELSE
    		\STATE $ \bar{v}_{randomly\,picked\,index}=\bar{v}_{randomly\,picked\,index}-1 $
    		\STATE $ d-=1 $
    		\ENDIF
    		\ENDWHILE
    		\ENDIF
    		\ENDFOR

    	\end{algorithmic}
    	\vspace{0.2in}
    	\hrule
    \end{center}
    \vspace{0.5in}
    \begin{center}
    	\hrule
    	\vspace{0.2in}
    	\captionof{algorithm}{Distribution Generator}
    	\hrule
    	\vspace{0.2in}
    	\begin{algorithmic}
    		\REQUIRE Total sum of the down scaled distribution, s\\
    		Standard deviation, $ \sigma $\\
    		\ENSURE Down scaled approximation over the length of the aggregate, $ \bar{v} $
    		\STATE $ \bar{v}=$Fit the decided distrubiton to the given down scaled time frame
    		\IF{elems in $ \bar{v}<0 $}
    		\STATE $ \bar{v}=\bar{v}+|min(\bar{v})| $
    		\ENDIF
    		\IF{elems in $ \bar{v}$ are not integer}
    		\STATE $ \bar{v}=round(\bar{v}) $
    		\ENDIF
    		\IF{$ s\neq \sum \bar{v} $}
    		\STATE d=$ s- \sum \bar{v} $
    		\WHILE{$ d\neq0 $}
    		\IF{$ d>0 $}
    		\STATE $ \bar{v}_{randomly\,picked\,index}=\bar{v}_{randomly\,picked\,index}+1 $
    		\STATE $ d-=1 $
    		\ELSE
    		\STATE $ \bar{v}_{randomly\,picked\,index}=\bar{v}_{randomly\,picked\,index}-1 $
    		\STATE $ d-=1 $
    		\ENDIF
    		\ENDWHILE
    		\ENDIF
    	\end{algorithmic}
    	\vspace{0.2in}
    	\hrule

    \end{center}

    The MKD algorithm is heavily dependent on the random selection of numbers that are prone to generate non-reproducible results. Thus seeding the random number generator is highly recommended to ensure reproducible results.\\

    The novelty of MKD algorithm is its consideration of the prior distribution as initialization and deploying the underlying distribution to generate synthesized downscaled data, which is non-negative and conserves the aggregated value of the given data.

    \section{Comparison of the Synthesized Data with the Real Data}\label{sec:realvssynth}

    To determine the accuracy of the MKD algorithm, we test the MKD algorithm against some real-world data. Here, we have taken 2020 COVID-19 data on infected individuals in Bangladesh and 2022 (January to July), Dengue data on infected individuals in Bangladesh. The aforementioned data are daily data of the number of newly infected individuals across the country. We aim to convert this data to monthly aggregate and feed the aggregated data to the algorithm to generate downscaled daily data; hence we can compare the accuracy of the synthetic daily data with respect to the actual daily data. To determine the accuracy of the approximation, we will use two error measures, and we will do component analysis on the real and synthetic data to see if the synthetic data can well approximate the underlying properties of the real data. In case of the component decomposition, we will use the additive model mentioned in \eqref{eq:additivemodel},
    \begin{equation}\label{eq:additivemodel}
    	y_i=Trend+Seasonality+Residual
    \end{equation}
    as the procured data has some zero values for which the multiplicative model mentioned in \eqref{eq:multiplicativemodel}

    \begin{equation}\label{eq:multiplicativemodel}
    	y_i=Trend\times Seasonality\times Residual
    \end{equation}

    is not suitable in this scenario.

        \subsection{Error Measures for Benchmark}

        To compare the result with the real world data we shall use two error terms that describes the overall error of the approximation. These are as follows:

        \begin{itemize}
        	\item \textbf{Root Mean Square Error:}\\
            The root-mean-square deviation (RMSD) or root-mean-square error (RMSE) is a commonly used measure of the discrepancies between the values (sample or population values) predicted by a model or estimator and the actual values. RMSD is the square root of the second sample moment of the discrepancies between anticipated and observed values, or the quadratic mean of these differences. When the computations are executed over the data set used for estimate, these deviations are known as residuals, and when they are computed out-of-sample, they are known as errors (or prediction errors). The RMSD aggregates the magnitudes of the mistakes in predictions for various pieces of data into a single metric of predictive ability. RMSD is a measure of accuracy used to assess the predicting losses of various models for a specific dataset and not between datasets because it is scale-dependent \cite{hyndman2006another}.\\

            RMSD is always positive, and a value of 0 would suggest a perfect fit to the data, which is nearly never attained in practice. A smaller RMSD is often preferable to a greater one. However, because the metric is dependent on the magnitude of the numbers used, comparisons between various kinds of data are invalid.\\

            The root square of the mean of squared mistakes is the RMSD. The influence of each inaccuracy on RMSD is proportional to the magnitude of the squared error; therefore, larger errors have an outsized effect on RMSD. As a result, the RMSD is extremely sensitive to outliers \cite{pontius2008components,willmott2006use}.\\

            Instead of the Root Mean Square Deviation, the Mean Absolute Error (MAE) has been suggested as a useful statistical tool by a number of scholars. The MAE has certain advantages over the RMSD when it comes to interpretability. The mean absolute error, abbreviated as MAE, is the average of the absolute values of the mistakes. The square root of the average of squared errors is more difficult to grasp than the MAE, which simplifies things considerably. In addition, the magnitude of a mistake has an effect on the MAE in direct proportion to its absolute value, whereas the RMSD does not follow this pattern at all \cite{pontius2008components}.\\

            RMSE can be defined using the following formula:

        			\begin{center}
                        RMSE=$ \sqrt{ \displaystyle \frac{\sum_{i=1}^N (x_i-\hat x_i)}{N} } $
                    \end{center}
            where, $ x_i $ is the actual data and $ \hat x_i $ is the predicted data.
        	\item \textbf{Mass Absolute Error:}\\
            In statistics, the term ``mean absolute error" (MAE) refers to a measurement of the errors that occur when matched observations expressing the same event are compared. Comparisons of what was predicted vs what was actually observed, subsequent time versus beginning time, and one technique of measurement versus an alternate technique of measurement are all examples of Y versus X. The mean absolute error (MAE) is determined by taking the sum of all absolute errors and dividing it by the total number of samples:\\
        		\begin{center}
                    	MAE=$ \displaystyle \frac{\sum_{i=1}^N |x_i -\hat x_i|}{N} $
                \end{center}
            Therefore, it is an arithmetic average of the absolute errors, which may be represented as $|e_{i}|=| x_{i}-\hat x_{i}| $, where $ \hat x_i $ represents the forecast and  $  x_i $ represents the actual value. It is important to keep in mind that different formulations could use relative frequencies as weight factors. The scale that is being used to measure the data is also used for the mean absolute error. Because this is what's known as a scale-dependent accuracy measure, it can't be used to compare series that have different scales because the comparisons wouldn't be valid \cite{hyndman2018forecasting}.  In time series analysis, the mean absolute error is a frequent way to quantify the accuracy of forecasts \cite{hyndman2006another}, occasionally leading to confusion with the more traditional definition of mean absolute deviation. There is, more generally speaking, the same confusion.\\

            One of the many methods that may be used to compare forecasts with the results that actually transpired is called the mean absolute error. The mean absolute scaled error, also known as MASE, and the mean squared error are two options that have a solid track record. The mean signed difference is one metric that does put emphasis on this, as opposed to the other measures, which all summarize performance in a fashion that disregards the direction of whether the forecast was made too high or too low.\\

            When it comes to fitting a prediction model with a chosen performance metric, the equivalent for mean absolute error is mean absolute deviations, and the least squares approach is related to the mean squared error.\\

            Although some academics report and interpret it that way, mean absolute error (MAE) and root-mean-square error (RMSE) are not the same concept. The MAE is conceptually simpler and also easier to perceive than the RMSE. It is just the average absolute vertical or horizontal distance between each point in a scatter plot and the Y=X line. In contrast, the RMSE is a measure of error that is more difficult to interpret. To phrase this another way, MAE refers to the average absolute difference that exists between X and Y. In addition, the contribution that each error makes to the MAE is weighted according to the absolute value of the error. This is in contrast to the RMSE, which involves quadrupling the differences; as a result, a few significant changes will have a higher impact on the RMSE than they will have on the MAE \cite{pontius2008components}.
        \end{itemize}

        Since many of the data points in the actual and synthesized  cases is popluated with $ 0 $ hence Mass Absolute Percentage Error (MAPE), and Scaled Mass Absolute Percentage Error (MAPE) are undefined in this scenario.

    	\subsection{Dengue}\label{subsec:dengue}

    	\subsubsection{Preprocessing and Result}
    	In case of this simulation, we took Bangladesh's 2022 daily Dengue infected data from January to July. To feed this data into the MKD algorithm, we convert the daily data to monthly aggregate as illustrated in Figure \ref{fig:monthlydengue},

    	\begin{center}
    		\includegraphics[width=\textwidth]{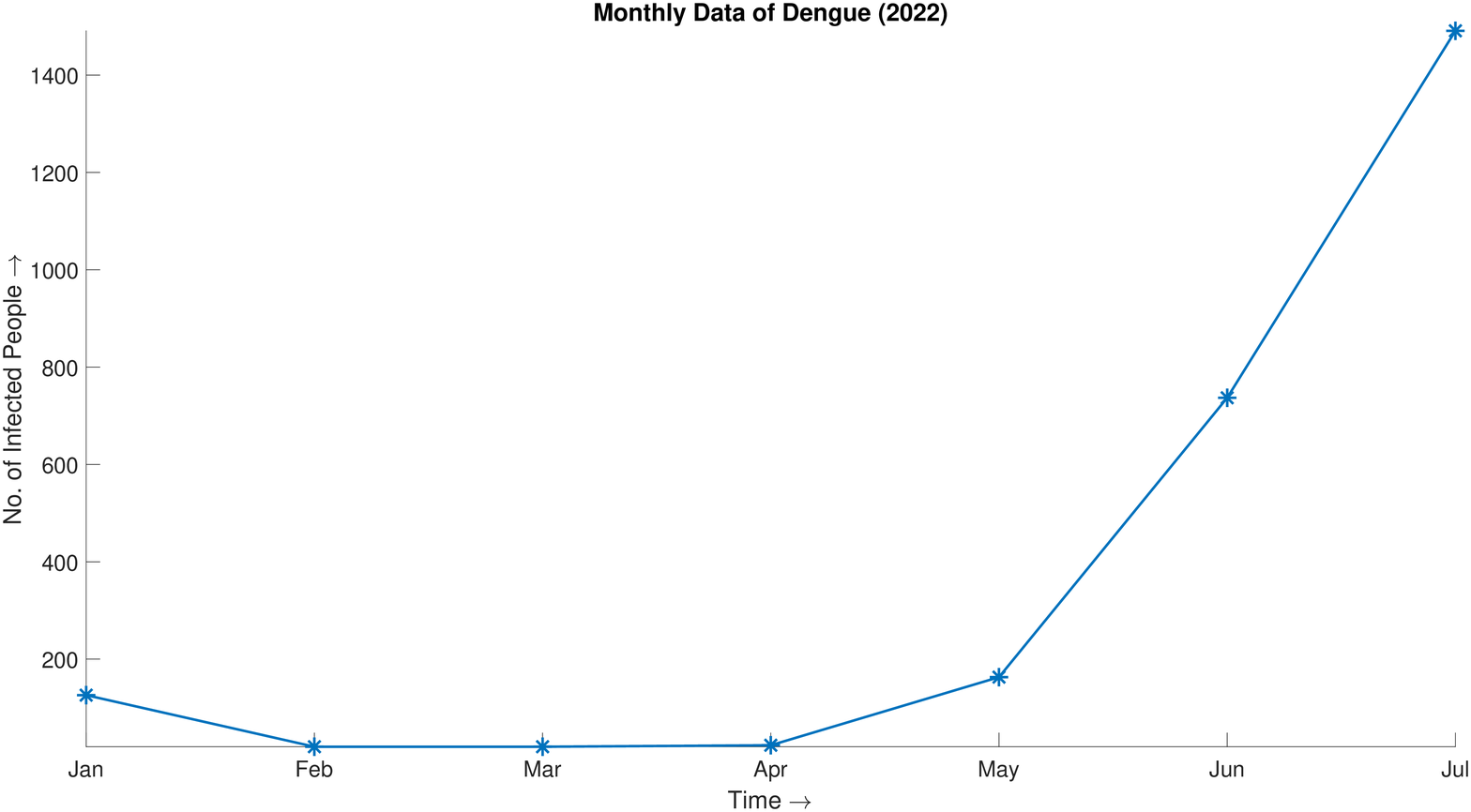}
    		\captionof{figure}{Monthly aggregate of 2022 Dengue data from January to July.}
    		\label{fig:monthlydengue}
    	\end{center}
    	We feed in this data considering,
    	\begin{itemize}

    		\item Initial standard deviation, $ \sigma_0=\displaystyle \frac{\sigma_{prior\,distribution}}{30}=\frac{556.6431703}{30}=18.55477234 $.

    		\item Over throw tolerance, $ \delta=0.6\times$ (Range of the initial distribution).

    		\item Iteration limit, $ n=100 $.

    		\item Radius of open interval, $ r=3 $.

    		\item Underlying distribution to be normal.
    	\end{itemize}
    	and generate the synthesized data. Figure \ref{fig:dailydenguesynthesized} illustrates the synthesized data, which can be said to be a good approximation of the actual given the aggregated prior distribution.\\

    	\begin{minipage}{0.5\textwidth}

    		\centering
    		\includegraphics[width=\textwidth]{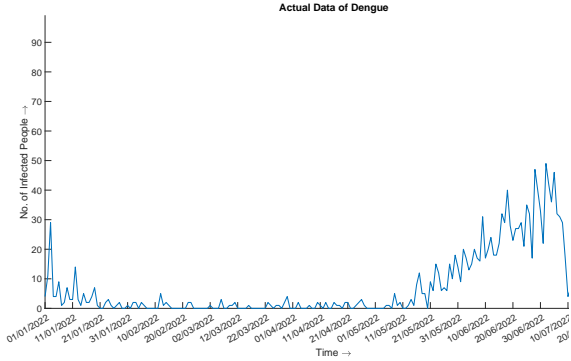}
    		\captionof{figure}{Daily number of infected cases of Dengue in 2022 from January to July collected from DGHS.}
    		\label{fig:dailydengueactul}
    	\end{minipage}%
    	\begin{minipage}{0.5\textwidth}
    		\centering

    		\includegraphics[width=\textwidth]{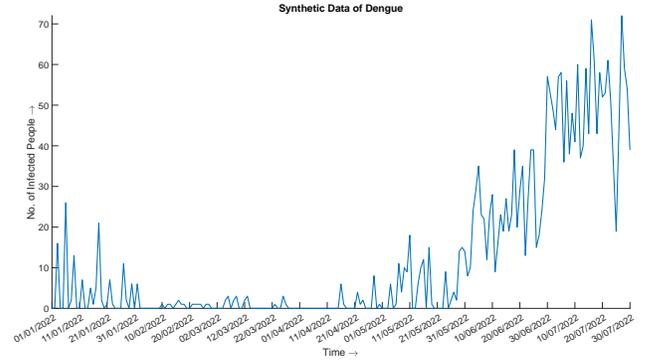}
    		\captionof{figure}{Synthesized daily number of infected cases of Dengue in 2022 from January to July.}
    		\label{fig:dailydenguesynthesized}

    	\end{minipage}

    	\subsubsection{Error Metrics and Statistical Measures}

    	The calculated error measures are:
    	\begin{itemize}
    		\item $ MAE=6.60664 $, which implies that the average error between the actual and synthesized data is $ 6.60664 $.

    		\item $ RMSE=12.64499 $ which implies that the standard deviation of the residuals/errors is $12.64499 $. The fact is well illustrated in Figure \ref{fig:syntheresidual}.
    	\end{itemize}
    	The error metric shows satisfactory results.
    	The following table validates if the synthesized data honours the aggregated sum of the prior distribution.
    	\begin{center}
    		\begin{tabular}{|c|c|>{\centering\arraybackslash}p{2.5cm}|>{\centering\arraybackslash}p{2.5cm}|>{\centering\arraybackslash}p{2.5cm}|}
    			\hline
    			\rowcolor{blue!20}\Centering\bfseries Month & \Centering\bfseries Actual & \Centering\bfseries Initial Distribution & \Centering\bfseries Overthrow Correction &\Centering\bfseries Volume Correction  \\ \hline
    			January & 126 & 126 & 119 & 126  \\ \hline
    			February & 20 & 20 & 27 & 20  \\ \hline
    			March & 20 & 20 & 20 & 20  \\ \hline
    			April & 23 & 23 & 32 & 23  \\ \hline
    			May & 163 & 163 & 206 & 163  \\ \hline
    			June & 737 & 737 & 733 & 737  \\ \hline
    			July & 1491 & 1491 & 1443 & 1491  \\ \hline
    			\bfseries Total & \textbf{2580} & \textbf{2580} & \textbf{2580} & \textbf{2580}  \\ \hline
    		\end{tabular}
    		\captionof{table}{This table illustrates that the synthetic data agrees with the monthly sum of the actual data}
    		\label{tab:voldengue2022}
    	\end{center}
    	The total number of cases in each scenario has been maintained equally. As discussed earlier, we can see that the initial distribution holds the monthly sum consistently, which gets disrupted in the overthrow correction phase and later corrected in the volume correction phase.\\

    	We shall now explore the basic statistical properties of the synthetic data with respect to the actual data.

    	\begin{table}[!ht]
    		\centering
    		\begin{tabular}{|c|r|r|}
    			\hline
    			\rowcolor{blue!20} \Centering\bfseries Measures &\Centering \bfseries Observed & \Centering\bfseries Synthesized  \\ \hline
    			Mean & 12.22748815 & 12.22748815  \\ \hline
    			Standard Deviation & 20.28993189 & 18.49672823  \\ \hline
    			Minimum & 0 & 0  \\ \hline
    			Lower Quartile(Q1) & 0 & 0  \\ \hline
    			Median & 2 & 1  \\ \hline
    			Upper Quartile(Q2) & 17 & 19  \\ \hline
    			Maximum & 99 & 72  \\ \hline
    		\end{tabular}
    		\caption{This table illustrates the comparison of the basic statistical measures of the synthesized data with respect to the actual data.}
    		\label{tab:statdengue2022}
    	\end{table}
    	It is to be noted that the mean of the synthesized data equates to that of the original data, although it was not plugged into the MKD algorithm in any manner. As previously discussed that $ \sigma_0 $ is a good approximation to the original $ \sigma $. All the rest of the measures are somewhat close, but the maximum varies by a lot. The maximum is hard to anticipate from the aggregated data; hence it is an avenue that demands further exploration.

    	\subsubsection{Component Decomposition and Comparison}

    	We now want to do component decomposition of both the actual and synthetic data based on the model mentioned in \eqref{eq:additivemodel}. However, component decomposition in no way is a benchmark for accuracy, but as MKD algorithm aims to improve the outcome of forecasting techniques which are highly influenced by the components within a time series data. Thus comparing these components can answer the question of whether the components-based characteristics of the original time series are present within the synthesized data.\\

    	In case of the trend component (Figure \ref{fig:actualtrend} and \ref{fig:synthetictrend} on appendix    \ref{app:fig}), both the actual and the synthesized data shows similar result and trend of the actual data have been well approximated by the trend of the synthesized data.\\

    	In case of the seasonality component (Figure \ref{fig:actualseason} and \ref{fig:syntheticseason} on appendix    \ref{app:fig}), both the actual and the synthesized data shows major weekly, and minor sub-weekly seasonality. The synthesized data's seasonality approximates the actual data's seasonality well.\\

    	In case of the residual component (Figure \ref{fig:actualresidual} and \ref{fig:syntheresidual} on appendix    \ref{app:fig}), both the actual and the synthesized data show a similar result, although the residual of the synthetic data may look a bit noisy at first glance but upon closer inspection, it is evident that the residual of the synthetic data shows less deviation from the standard value in comparison to the actual data. The actual data's residual has been well approximated by the synthesized data's residual.\\

    	The key takeaway from the discussion as mentioned earlier, is that the MKD algorithm could generate an excellent approximation of the dengue data from the monthly aggregated data based on some statistical properties of the prior distribution. We shall also test MKD algorithm's efficacy in another epidemiological scenario in the following section.

    	\subsection{COVID-19}\label{subsec:COVID19}

    	\subsubsection{Preprocessing and Result}

    	In case of this simulation, we took Bangladesh's 2020 daily COVID-19 infected data from March to December. To feed this data into the MKD algorithm, we convert the daily data to monthly aggregate as illustrated in Figure \ref{fig:monthlyCOVID},

    	\begin{center}
    		\includegraphics[width=\textwidth]{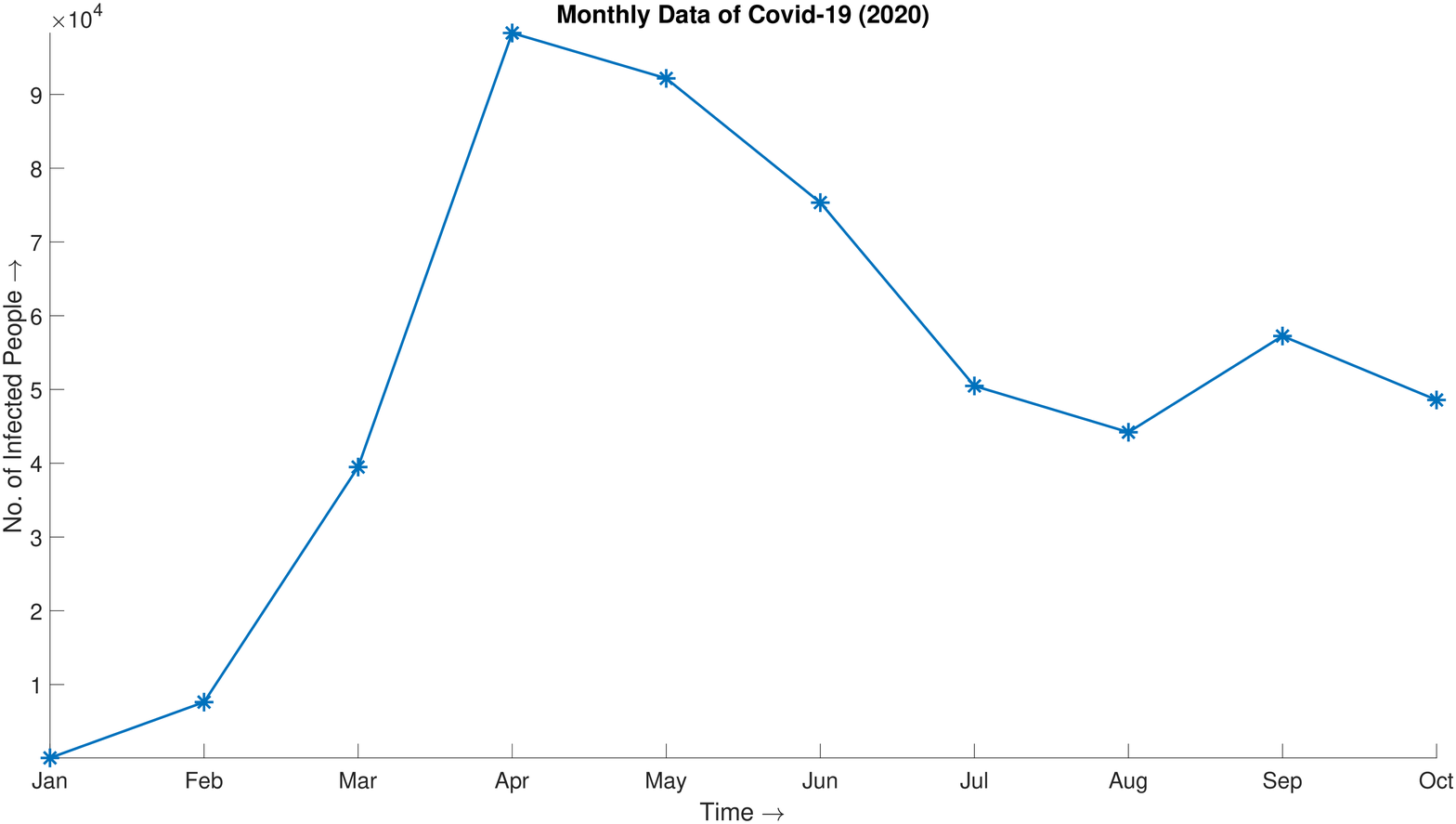}
    		\captionof{figure}{Monthly aggregate of 2020 COVID-19 infected data of Bangladesh from March to December.}
    		\label{fig:monthlyCOVID}
    	\end{center}
    	We feed in this data considering,
    	\begin{itemize}

    		\item Initial standard deviation, $ \sigma_0=\displaystyle \frac{\sigma_{prior\,distribution}}{30}=\frac{32021.87439}{30}=1067.395813 $.

    		\item Over throw tolerance, $ \delta=0.2\times$ (Range of the initial distribution).

    		\item Iteration limit, $ n=100 $.

    		\item Radius of open interval, $ r=3 $.

    		\item Underlying distribution to be normal.
    	\end{itemize}
    	and generate the synthesized data. Figure \ref{fig:dailyCOVIDsynthesized} illustrates the synthesized data, which can be said to be a good approximation of the actual given the aggregated prior distribution.\\

    	\begin{minipage}{0.5\textwidth}

    		\centering
    		\includegraphics[width=\textwidth]{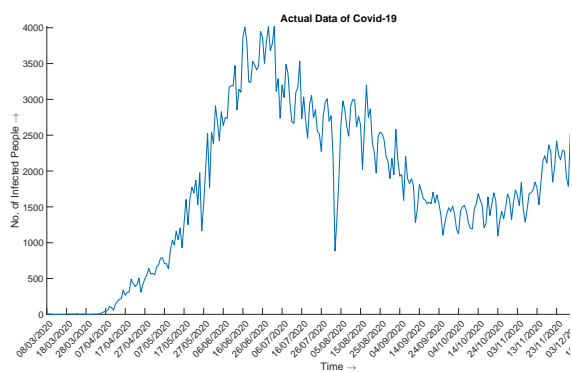}
    		\captionof{figure}{Daily number of infected cases of COVID-19 in 2020 from March to December collected from DGHS.}
    		\label{fig:dailyCOVIDactual}
    	\end{minipage}%
    	\begin{minipage}{0.5\textwidth}
    		\centering

    		\includegraphics[width=\textwidth]{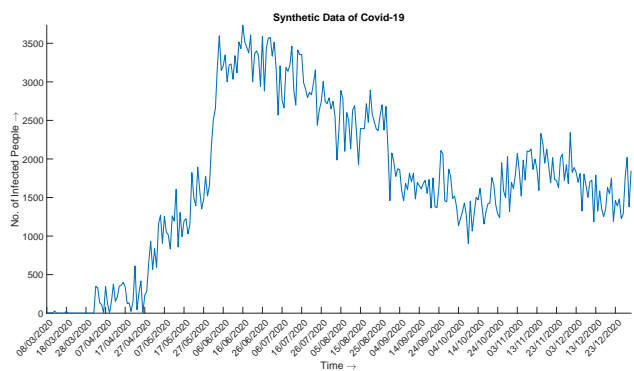}
    		\captionof{figure}{Synthesized daily number of infected cases of COVID-19 in 2020 from March to December.}
    		\label{fig:dailyCOVIDsynthesized}

    	\end{minipage}
    	%

    	\subsubsection{Error Metrics and Statistical Measures}
    	The calculated error measures are:
    	\begin{itemize}
    		\item $ MAE=257.41806 $, which implies that the average error between the actual and synthesized data is $ 257.41806 $, which is reasonable considering the mean of the data is $ 1717.424749 $.

    		\item $ RMSE=346.6241$, which implies that the standard deviation of the residuals/errors is $346.6241 $. The fact is well illustrated in Figure \ref{fig:syntheresidual1}.
    	\end{itemize}
    	it is to be noted that the error term of this scenario must not be compared with the error term of the previous case as they are of varying scale. Compared to the scale of the data, the error metric shows satisfactory results.
    	The following table validates if the synthesized data honours the aggregated sum of the prior distribution.
    	\begin{center}
    		\begin{tabular}{|c|c|>{\centering\arraybackslash}p{2.5cm}|>{\centering\arraybackslash}p{2.5cm}|>{\centering\arraybackslash}p{2.5cm}|}
    			\hline
    			\rowcolor{blue!20}\Centering\bfseries Month & \Centering\bfseries Actual & \Centering\bfseries Initial Distribution & \Centering\bfseries Overthrow Correction & \Centering\bfseries Volume Correction \\ \hline
    			March                  &                        51 &                                      51 &                                      51 &                                   51 \\ \hline
    			April                   &                         7616 &                                       7616 &                                       9226 &                                    7616 \\ \hline
    			May                    &                         39486 &                                       39486 &                                       41261 &                                    39486 \\ \hline
    			June                    &                         98330 &                                       98330 &                                       94075 &                                    98330 \\ \hline
    			July                     &                        92178 &                                      92178 &                                      92115 &                                   92178 \\ \hline
    			August                     &                        75335 &                                      75335 &                                      75605 &                                   75335 \\ \hline
    			September                    &                       50483 &                                     50483 &                                     50766 &                                  50483 \\ \hline
    			October                     &                       44205 &                                     44205 &                                     45126 &                                  44205 \\ \hline
    			November                     &                       57248 &                                     57248 &                                     55805 &                                  57248 \\ \hline
    			December                     &                       48578 &                                     48578 &                                     49480 &                                  48578 \\ \hline
    			\bfseries Total               &              \textbf{513510} &                            \textbf{513510} &                            \textbf{513510} &                         \textbf{513510} \\ \hline
    		\end{tabular}
    		\captionof{table}{This table illustrates that the synthetic data agrees with the monthly sum of the actual data.}
    		\label{tab:volCOVID2020}
    	\end{center}
    	The total number of cases in each scenario has been maintained equally. As discussed earlier, we can see that the initial distribution holds the monthly sum consistently, which gets a little disrupted in the overthrow correction phase and is later on corrected in the volume correction phase.\\

    	We shall now explore the basic statistical properties of the synthetic data with respect to the actual data.
    	\begin{center}
    		\centering
    		\begin{tabular}{|c|r|r|}
    			\hline
    			\rowcolor{blue!20} \Centering\bfseries Measures &\Centering \bfseries Observed & \Centering\bfseries Synthesized  \\ \hline
    			Count & 299 & 299  \\ \hline
    			Mean & 1717.424749 & 1717.424749  \\ \hline
    			Standard Deviation & 1044.457258 & 1007.554237  \\ \hline
    			Minimum & 0 & 0  \\ \hline
    			Lower Quartile(Q1) & 1115.5 & 1225  \\ \hline
    			Median & 1666 & 1696  \\ \hline
    			Upper Quartile(Q2) & 2521.5 & 2481.5  \\ \hline
    			Maximum & 4019 & 3735  \\ \hline
    		\end{tabular}
    		\captionof{table}{This table illustrates the comparison of the basic statistical measures of the synthesized data with respect to the actual data.}
    		\label{tab:statCOVID2020}
    	\end{center}

    	It is to be noted that the mean of the synthesized data equates to that of the original data, although it was not plugged into the MKD algorithm in any manner. As previously discussed that $ \sigma_0 $ is a good approximation to the original $ \sigma $. All the rest of the measures are somewhat close, but the maximum varies by a lot. The maximum is hard to anticipate from the aggregated data; hence it is an avenue that demands further exploration.

    	\subsubsection{Component Decomposition and Comparison}

    	We now want to do component decomposition of both the actual and synthetic data based on the model mentioned in \eqref{eq:additivemodel}. However, component decomposition in no way is a benchmark for accuracy, but as MKD algorithm aims to improve the outcome of forecasting techniques which are highly influenced by the components within a time series data. Thus, comparing these components can answer the question of whether the original time series's components-based characteristics are present in the synthesized data.\\

    	In case of the trend component (Figure \ref{fig:actualtrend1} and \ref{fig:synthetictrend1} on appendix    \ref{app:fig}), both the actual and the synthesized data shows similar result and trend of the actual data have been well approximated by the trend of the synthesized data.\\

    	In case of the seasonality component (Figure \ref{fig:actualseason1} and \ref{fig:syntheticseason1} on appendix    \ref{app:fig}), both the actual and the synthesized data shows major weekly seasonality. The seasonality of the synthesized data has well approximated the seasonality of the actual data.\\

    	In case of the residual component (Figure \ref{fig:actualresidual1} and \ref{fig:syntheresidual1} on appendix    \ref{app:fig}), both the actual and the synthesized data shows a similar result, although the residual of the synthetic data may look a bit noisy at first glance but upon closer inspection, it is evident that the residual of the synthetic data shows less deviation from the standard value in comparison to the actual data. The residual of the synthesized data has well approximated the residual of the actual data.\\

    	The key takeaway from the aforementioned discussion is that the algorithm could generate an excellent approximation of the COVID-19 data from the monthly aggregated data based on some statistical properties of the prior distribution. We shall also test MKD algorithm's efficacy in a forecasting scenario in the following section.

    	\section{Improvements in Forecasting Accuracy}\label{sec:forecast}

    	In this section, we shall forecast the Dengue infection case in Bangladesh using statistical forecasting tools. The use of statistical modelling is one of the helpful ways that may be utilized for the forecasting of dengue outbreaks \cite{husin2008modeling,wong2015factors}. Previous research carried out in China \cite{lu2009time}, India \cite{bhatnagar2012forecasting}, Thailand \cite{wongkoon2007predicting}, West Indies \cite{gharbi2011time}, Colombia \cite{torres2014fuzzy}, and Australia \cite{hu2010dengue} on infectious diseases made substantial use of the time series technique in the field of epidemiologic research on infectious diseases \cite{hu2010dengue}.
    	A number of earlier research looked at the Autoregressive Integrated Moving Average (ARIMA) model as a potential tool for use in forecasting \cite{abdulla2015forecasting,hossain2016forecasting,hossain2015jute,hossain2016forecasting,hossain2015forecasting,hossain2015forecasting,hossain2016forecasting,hossian2015time}.
    	In addition, the ARIMA models have seen widespread use for dengue forecasting \cite{Earnest2012,eastin2014intra,hossain2015forecasting,wu2007weather}.
    	When establishing statistical forecasting models, these are frequently paired with Seasonal Auto-regressive Integrated Moving Average (SARIMA) models, which have proven to be suitable for assessing time series data with ordinary or seasonal patterns \cite{bhatnagar2012forecasting,gharbi2011time,hu2010dengue,luz2008time,martinez2011sarima}. It is likely that developing a dengue incidence forecasting model based on knowledge from previous outbreaks and environment variables might be an extremely helpful tool for anticipating the severity and frequency of potential epidemics.\\

    	ARIMA is a well-known model in statistics that is predominantly used to forecast and analyze time series data \cite{brownlee2017introduction}. Auto Regression of order p can be defined as
    	\begin{equation}\label{eq:ARmodel}
    		Y_t=e_t+\sum_{i=1}^{p} \alpha_iY_{t-i}
    	\end{equation}
    	where $ {e_t} $ are white noises of mean $ 0 $ and variance $ \sigma_e^2 $.\\
    	The Moving Average (MA) of order q is defined as,
    	\begin{equation}\label{eq:MAmodel}
    		\hat{Y}_t=e_t+\sum_{i=1}^{q} \beta_ie_{t-i}
    	\end{equation}

    	ARMA model in theory is formed in unison of \eqref{eq:ARmodel} and \eqref{eq:MAmodel}. Hence, an ARMA model of order $ (p,\,q) $ is defined
    	\begin{equation}\label{eq:ARMAmodel}
    		\hat{Y}_t=e_t+\sum_{i=1}^{p} \alpha_iY_{t-i} +\sum_{i=1}^{q} \beta_ie_{t-i}
    	\end{equation}
    	where the p and q are the corresponding order of the AR and MA. Development of the ARMA model for non-stationary time series is the Box-Jenkins model, also known as the ARIMA model, which integrates AR and MA with successive difference/lag operator, $ \triangledown^d $. Hence, an ARIMA model of order $ (p,\,d,\,q) $ is defined
    	\begin{equation}\label{eq:ARIMAmodel}
    		\hat{Z}_t=e_t+\sum_{i=1}^{p} \alpha_iZ_{t-i}+\sum_{i=1}^{q} \beta_ie_{t-i}
    	\end{equation}
    	where p, q has the previously mentioned definition, and $ d $ is the order of nonseasonal successive difference required to make the time series stationary i.e.
    	\begin{itemize}
    		\item If d=1 then $ Z_t=\triangledown Y_t=Y_t-Y_{t-1} $
    		\item If d=2 then $ Z_t=\triangledown^2 Y_t=(Y_t-Y_{t-1})-(Y_{t-1}-Y_{t-2})=Y_t-2Y_{t-1}+Y_{t-2} $
    		\item If d=3 then $ Z_t=\triangledown^3 Y_t=( Y_t-2Y_{t-1}+Y_{t-2} )-(Y_{t-1}-2Y_{t-2}+Y_{t-3})=Y_t-3Y_{t-1}+3Y_{t-2}-Y_{t-3}$
    		\item and so on.
    	\end{itemize}

    	The idea of seasonality using the Fourier coefficient naming Fourier ARIMA model was introduced by \cite{nachane2008forecasting}.

    	\begin{equation}\label{eq:fourierARIMA}
    		Z_t=\delta_0+\sum_{i=1}^p \alpha_l Z_{t-i}+\sum_{j=1}^q \beta_J e_{t-j}+
    		\sum_{k=1}^r\left[a_k \sin \left(\omega_k t\right)+b_k \cos \left(\omega_k t\right)\right] Z_{t-m}+e_t
    	\end{equation}

    	where, $ \delta_0 $ is the constant term and $ \omega_k $ is the periodicity of the data.\\

    	We aim to forecast the monthly data and the synthesized daily data using the aforementioned forecasting techniques and compare the accuracy of the forecast based on error measures. We use SARIMA and Fourier-ARIMA model to forecast the monthly and synthesized data respectively. The model in each case is chosen based on the lowest value of Akaike's Information Criterion (AIC),  Akaike's Information Criterion correction (AICc), and Bayesian Information Criterion (BIC).

    	\subsection{Model Selection Method}
    	Box-Jenkins method is a generalized model selection pathway which works for time series irrespective of its stationarity or seasonality. The method is illustrated in Figure \ref{fig:box_jenkin_method}.

    	\begin{center}
    		\includegraphics[scale=0.5]{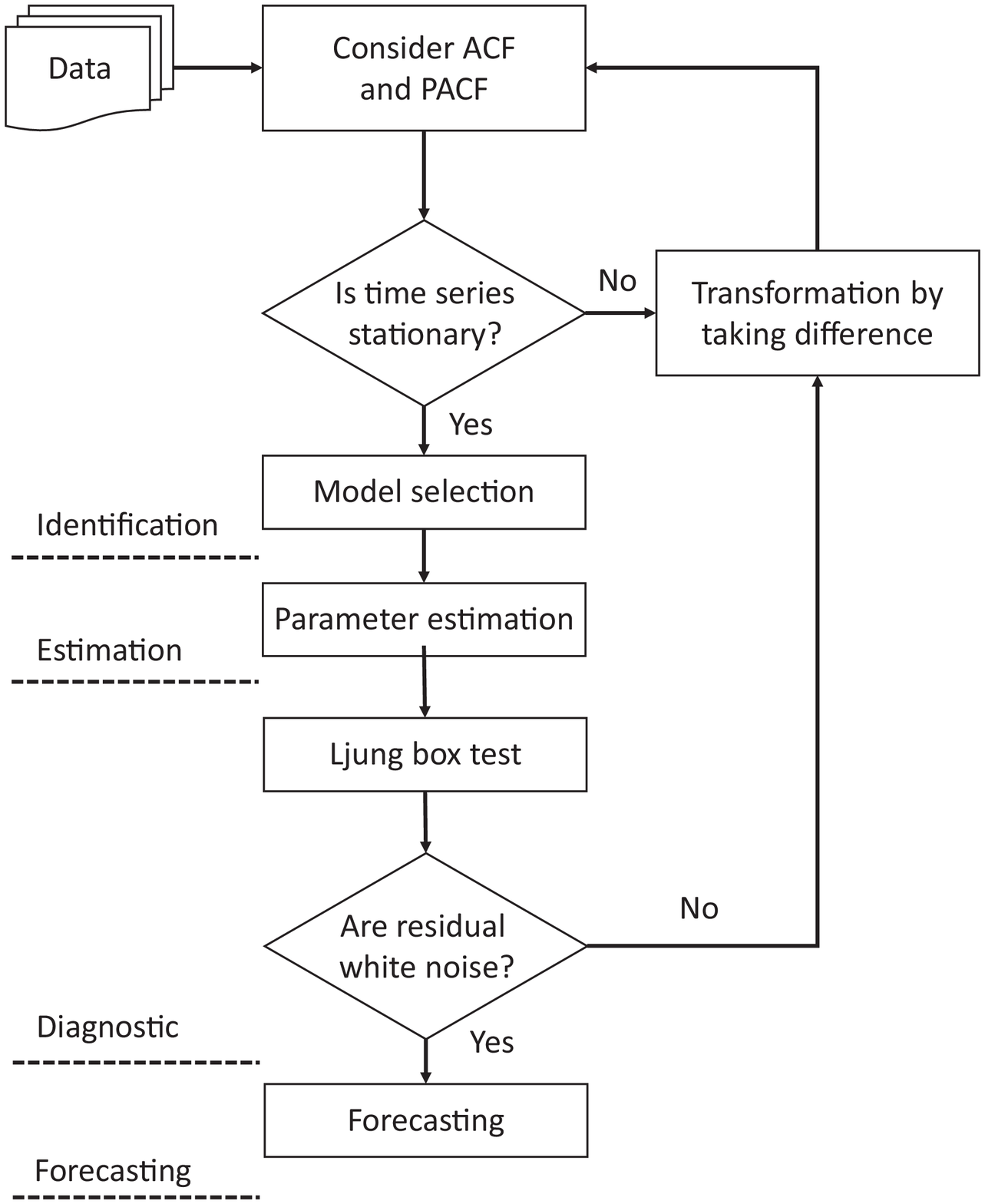}
    		\captionof{figure}{Flow chart of Box-Jenkin's Method.}
    		\label{fig:box_jenkin_method}
    	\end{center}

    	\subsection{Error Measures}
    	The error measures for comparison is Mean Absolute Scaled Error(MASE) which is defined as
    	$$ MASE=\frac{\frac{1}{n} \sum_{i=1}^n\left|Y_i-\hat{Y_i} \right|}{\frac{1}{T-m} \sum_{t=m+1}^T\left|Y_t-Y_{t-m}\right|} $$
    	We used this metric as it is scale-independent; hence is perfect for comparison. We also could have taken MAPE as a metric, but MAPE is undefined for such cases as the data is populated with zero values. We also use RMSE and MAE to gauge the error in the forecast.

    	\subsection{Forecast on the Aggregated Data}
    	The actual data is monthly Dengue infection data of Bangladesh from 2010 to July 2022. Following Box-Jenkin's method, we firstly check for the stationarity of the data based on the Augmented Dicky Fuller (ADF) test. ADF test returns the value of -4.7906 with p-value = 0.01, which implies that the data is stationary.\\

    	We run multiple SARIMA models and calculate their AIC, AICc and BIC and the best model is chosen based on the minimum value of the criterion. We present 5 of the top results in table \ref{tab:monthlymodelselection}.

    	\begin{center}
    		\begin{tabular}{|c|c|c|c|}
    			\hline
    			\rowcolor{blue!20} Model & AIC & AICc & BIC \\\hline
    			SARIMA$ (1, \,0, \,0)(0, \,1, \,1)_{12} $ & 2603.57  & 2603.76   & 2612.22\\ \hline

    			SARIMA$ (1, \,0, \,0)(0, \,1, \,2)_{12} $ & 2506.58   & 2506.91   & 2517.96\\ \hline

    			SARIMA$ (1, \,0, \,0)(1, \,1, \,1)_{12} $ & 2507.12   & 2507.45   & 2518.49\\ \hline

    			SARIMA$ (1, \,0, \,1)(0, \,1, \,1)_{12} $ & 2510.07   & 2510.39   & 2521.44\\ \hline

    			SARIMA$ (2, \,0, \,0)(0, \,1, \,1)_{12} $ & 2510.13    & 2510.45    & 2521.5\\ \hline
    		\end{tabular}
    		\captionof{table}{Selection of Best model based on criteria.}
    		\label{tab:monthlymodelselection}
    	\end{center}

    	Here, the best model to use is SARIMA $ (1, \,0, \,0)(0, \,1, \,1)_{12} $. We fit the given model, which gives us the following coefficients:
    	\begin{center}
    		\begin{tabular}{r|rrr}
    			& ar1 & ar2 & sma1 \\ \hline

    			& 0.6000 & -0.0919 & -0.8324 \\

    			\text { S.E. } & 0.0879 & 0.0877 & 0.0948

    		\end{tabular}
    		\captionof{table}{Coefficients of SARIMA $ (1, \,0, \,0)(0, \,1, \,1)_{12} $ model to fit and forecast actual monthly data of Dengue infection in Bangladesh from 2010 to July, 2022. Here, ar implies autoregressive, ma implies moving average, SMA implies seasonal moving average, and the trailing number enumerates their coefficient ordering. SE implies the standard error of the mean.}
    	\end{center}

    	To check the goodness of fit of the model, we use the Ljung box test, which returns the p-value = $ 0.9998 > 0.05 $, i.e. we accept the null hypothesis: \textit{``The model does not show lack ness of fit/ the residuals are not autocorrelated/ the residuals are random white noise.''}\\

    	Given everything in place, we forecast the infection for the rest of 2023, i.e. from August to December. The forecast is illustrated in the given figure.

    	\begin{center}
    		\includegraphics[width=\textwidth]{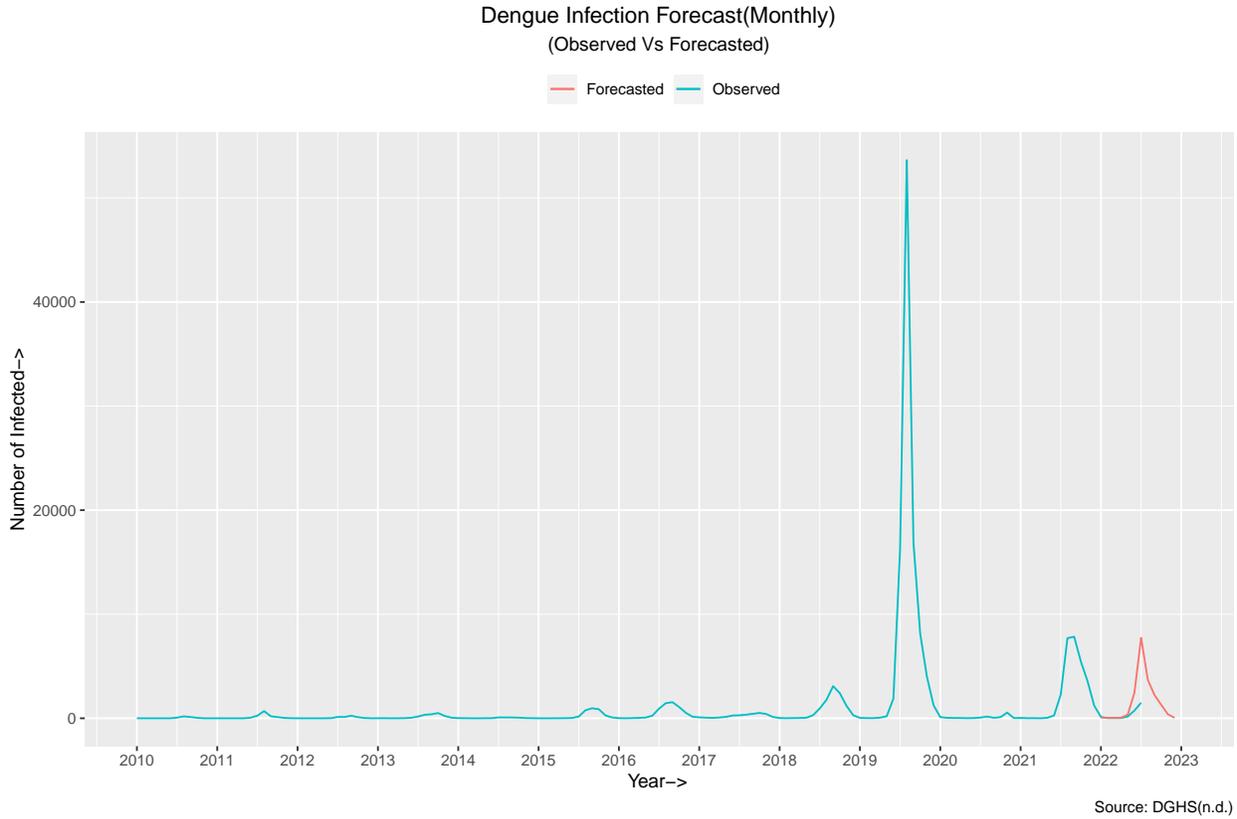}
    		\captionof{figure}{The figure illustrates the forecast generated by SARIMA $ (1, \,0, \,0)(0, \,1, \,1)_{12} $ from actual aggregated data.}
    	\end{center}

    	To validate the goodness of the fit, we can analyze the model residual, illustrated in Figure \ref{fig:monthlydata residual}. Here, the top graph is that of the residual with the timeline of the original data. The bottom left graph represents the Autocorrelation Function (ACF) with respect to the lag of the data. Almost all the values are within the significance e level, and the bottom right figure shows the distribution of the model's residuals. It implies that the residuals are distributed normally with zero mean.

    	\begin{center}
    		\includegraphics[width=\textwidth ]{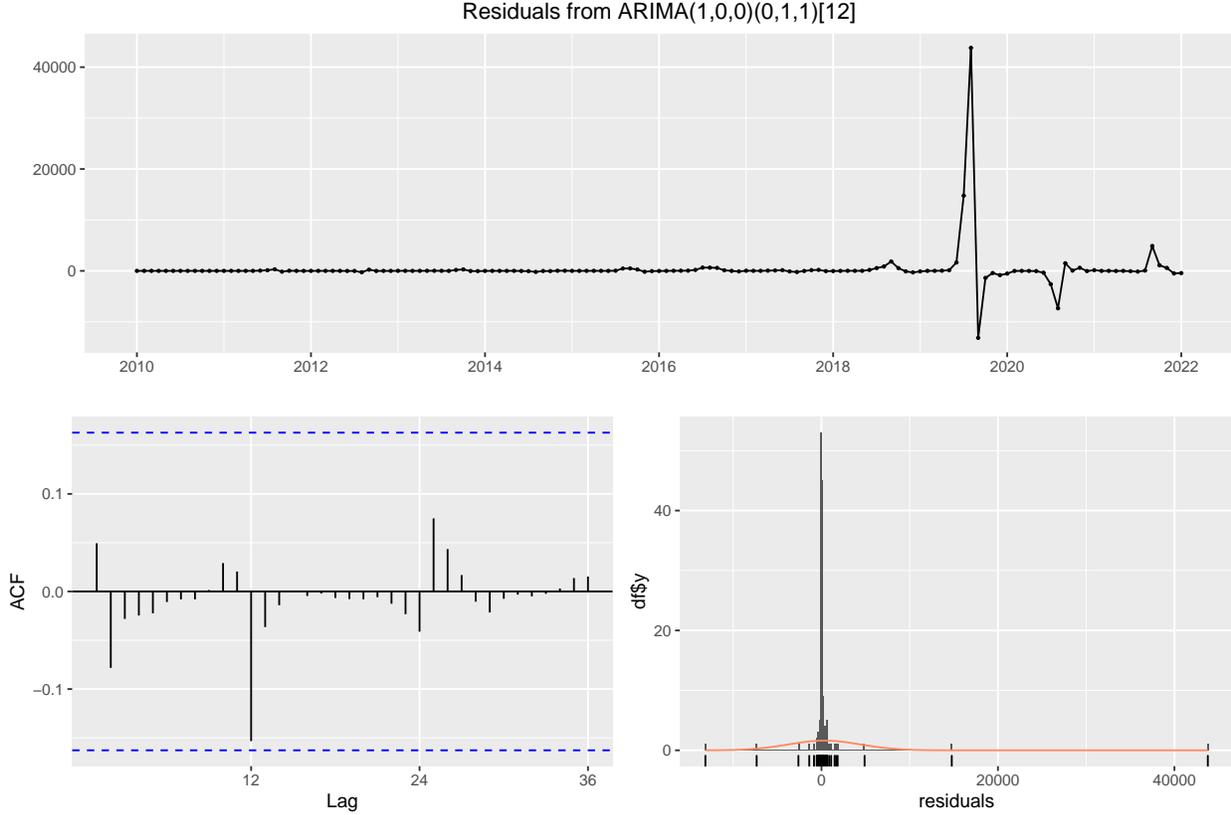}
    		\captionof{figure}{Residual of the SARIMA $ (1, \,0, \,0)(0, \,1, \,1)_{12} $.}
    		\label{fig:monthlydata residual}
    	\end{center}

    	To calculate the accuracy of the given forecast, we calculate the aforementioned error measures.

    	\begin{center}
    		\begin{tabular}{|c|c|c|c|}
    			\hline
    			\rowcolor{blue!20}Data   &    RMSE    &    MAE     &    MASE     \\ \hline
    			Monthly & $4092.712$ & $753.6765$ & $0.409654$  \\ \hline
    		\end{tabular}
    		\captionof{table}{Error measures for the forecast of the SARIMA $ (1, \,0, \,0)(0, \,1, \,1)_{12} $ of the actual aggregated data.}
    		\label{tab:actual_data_error}
    	\end{center}

    	The error measures are acceptable given the magnitude of the data, but there is room for improvement shall be demonstrated in the following subsection.

    	\subsection{Forecast on the Synthesized Data}
    	The synthesized data is daily Dengue infection data of Bangladesh from 2010 to July 2022. Following Box-Jenkin's method, we firstly check for the stationarity of the data based on the Augmented Dicky Fuller (ADF) test. ADF test returns the value of -6.6531 with p-value = 0.01, which implies that the data is stationary.\\

    	We run multiple Fourier ARIMA models and calculate their AIC, AICc and BIC. The best model is chosen based on the minimum value of the criterion. We present 5 of the top results in table \ref{tab:syntheticmodelselection}. Here in each case of Fourier transformation, we used one pair of trigonometric terms where each pair is comprised of a sine and a cosine term as defined in \eqref{eq:fourierARIMA} and the periodicity of the Fourier term is used to be 365.25. Prior to this we have used box cox transformation of $ \lambda = 0.49 $

    	\begin{center}
    		\begin{tabular}{|c|c|c|c|}
    			\hline
    			\rowcolor{blue!20} Model &   AIC    &   AICc   &   BIC    \\ \hline
    			ARIMA(7,0,7)       & 21711.25 & 21711.4  & 21827.03 \\ \hline
    			ARIMA(5,0,0)       & 21819.04 & 21819.08 & 21876.94 \\ \hline
    			ARIMA(3,0,0)       & 22147.88 & 22147.9  & 22192.91 \\ \hline
    			ARIMA(2,0,0)       & 22527.02 & 22527.04 & 22565.61 \\ \hline
    			ARIMA(1,0,0)       & 23476.24 & 23476.25 & 23508.4  \\ \hline
    			ARIMA(0,0,0)       & 33245.98 & 33271.71 & 33271.71 \\ \hline
    		\end{tabular}
    		\captionof{table}{Selection of Best model based on criteria.}
    		\label{tab:syntheticmodelselection}
    	\end{center}

    	Here, the best model to use is ARIMA(7,0,7). We fit the given model, which gives us the following coefficients:
    	\begin{center}
    		\begin{tabular}{|c|cccccccc|}
    			\hline
    			&     ar1 &     ar2 &     ar3 &     ar4 &    ar5 &     ar6 &       ar7 &    ma1   \\ \hline
    			& -0.5273 &  0.3109 &  1.2946 &  1.0562 & 0.2775 & -0.6222 &   -0.7940 & 0.8055   \\
    			S.E. &  0.0513 &  0.0310 &  0.0419 &  0.0755 & 0.0323 &  0.0353 &    0.0488 & 0.0471   \\ \hline
    			&     ma2 &     ma3 &     ma4 &     ma5 &    ma6 &     ma7 & intercept & s1-365   \\ \hline
    			&  0.0718 & -1.0032 & -1.1256 & -0.5327 & 0.3051 &  0.6454 &    3.3498 & 6.6197   \\
    			S.E. &  0.0340 &  0.0365 &  0.0602 &  0.0321 & 0.0356 &  0.0303 &    1.4789 & 1.6859   \\ \hline
    			&  c1-365 &         &         &         &        &         &           &          \\ \hline
    			& -0.7430 &         &         &         &        &         &           &          \\
    			S.E. &  1.6857 &         &         &         &        &         &           &         \\ \hline
    	\end{tabular}
    		\captionof{table}{Coefficients of ARIMA(7,0,7) model to fit and forecast actual monthly data of Dengue infection in Bangladesh from 2010 to July 2022. Here, ar implies auto-regressive, ma implies the moving average, s and c represent the coefficient of the sine and cosine of Fourier term, intercept implies the constant term, and the trailing number enumerates their coefficient ordering. SE implies the standard error of the mean.}
    			\label{tab:syn}
    	\end{center}

    	To check the goodness of fit of the model, we use the Ljung box test, which returns the p-value = $ 0.07749 > 0.05 $, i.e. we accept the null hypothesis: \textit{``The model does not show lack ness of fit/ the residuals are not autocorrelated/ the residuals are random white noise''}.\\

    	Given everything in place, we forecast the infection for the rest of 2023, i.e. from August to December. The forecast is illustrated in the given figure.

    	\begin{center}
    		\includegraphics[width=\textwidth]{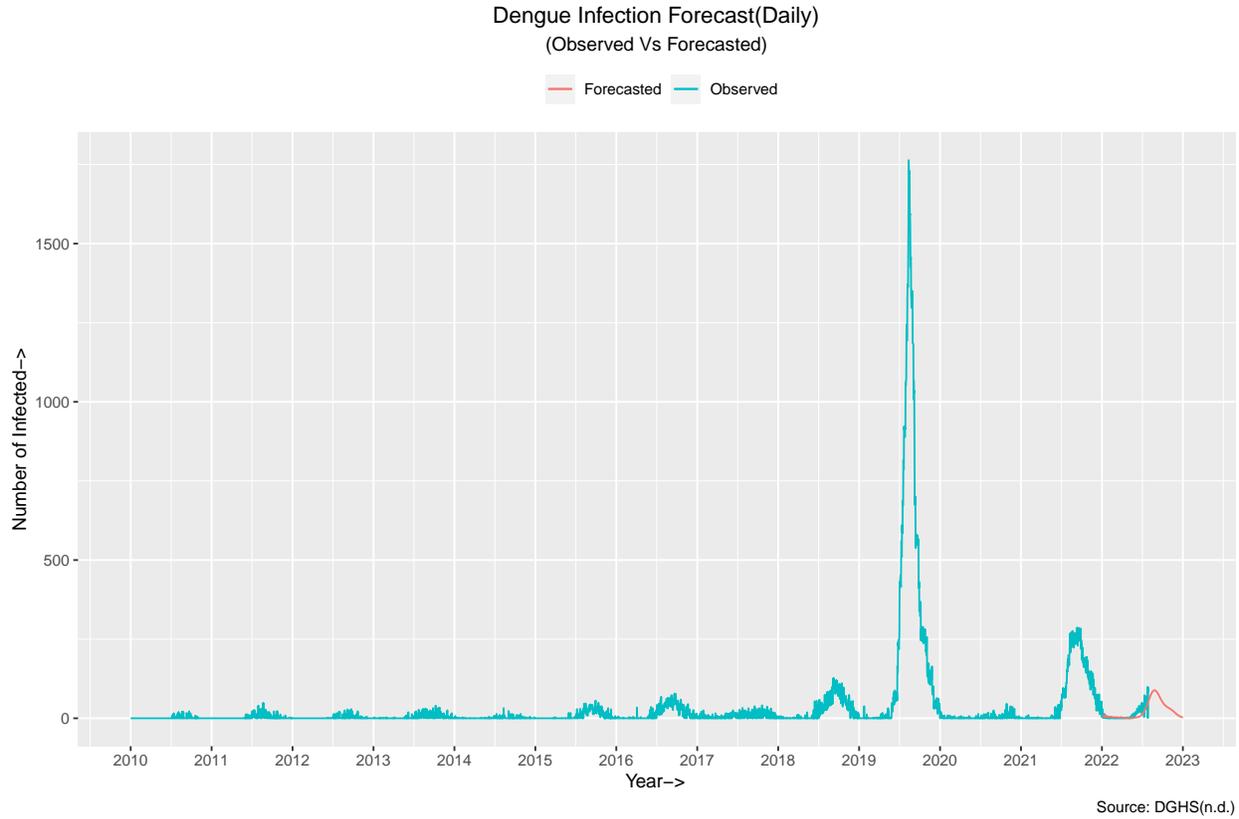}
    		\captionof{figure}{The figure illustrates the forecast generated by ARIMA(7,0,7) from actual aggregated data.}
    	\end{center}

    	To validate the goodness of the fit, we can analyze the model residual, illustrated in Figure \ref{fig:synthetic_data_residual}. Here, the top graph is that of the residual with the timeline of the original data. The bottom left graph represents the Autocorrelation Function (ACF) with respect to the lag of the data. Almost all the values are within the significance e level, and the bottom right figure shows the distribution of the model's residuals. It implies that the residuals are distributed normally with zero mean.

    	\begin{center}
    		\includegraphics[width=\textwidth ]{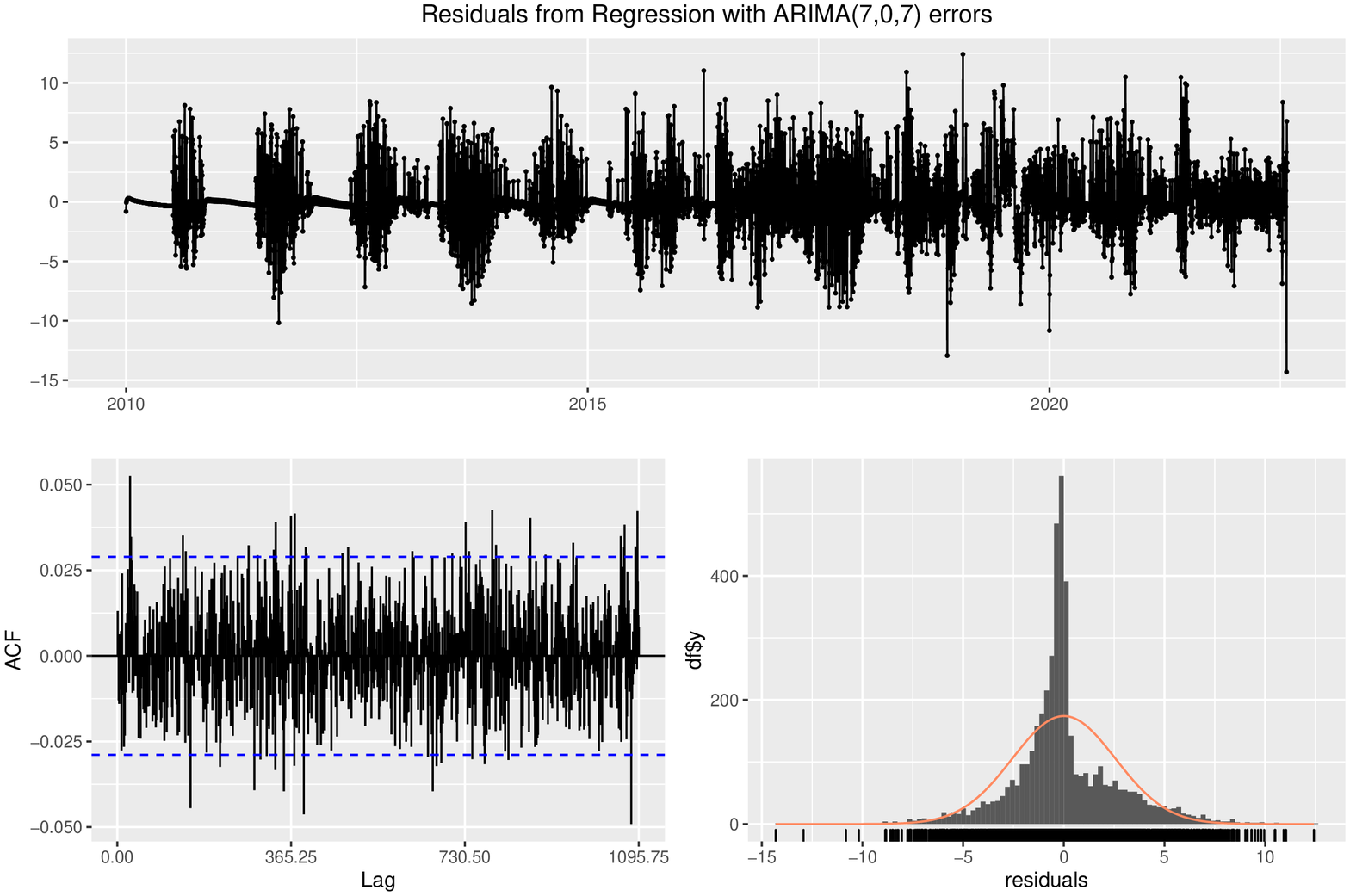}
    		\captionof{figure}{Residual of the ARIMA $ (7, \,0, \,7) $.}
    		\label{fig:synthetic_data_residual}
    	\end{center}

    	To calculate the accuracy of the given forecast, we calculate the aforementioned error measures.

    	\begin{center}
    		\begin{tabular}{|c|c|c|c|}
    			\hline
    			\rowcolor{blue!20}Data &    RMSE    &    MAE     &    MASE    \\ \hline
    			Daily         & $18.71255$ & $6.593062$ & $0.1115845$ \\ \hline
    		\end{tabular}
    		\captionof{table}{Error measures for the forecast of the ARIMA $ (7, \,0, \,7) $ of the synthetic daily data.}
    		\label{tab:synth_data_error}
    	\end{center}

    	The error measures are acceptable, given the magnitude of the data. In comparison to the error measures of   the actual data illustrated in table \ref{tab:actual_data_error}, we can see improvement in the table \ref{tab:synth_data_error}. Comparing the MASE term of the two tables shows about four times improvement in the forecast accuracy using the synthetic data over actual data.

    	\section{Conclusion}\label{sec:conclusion}
    	Downscaling algorithm has been predominantly used in geology to facilitate outputs of the prevalent models in the field. Very few applications have been made in epidemiology, and most of the application is spatial downscaling. This paper contributes by proposing a parametric, probabilistic one-dimensional downscaling algorithm using aggregated data in the field of epidemiology that facilitates existing forecasting tools box to generate better forecasts than the aggregated data. The MKD algorithm is by no means by construction applies to only epidemiological data but is tuned to work with non-negative, integer data. Deduction of particular conditioning on the initial data synthesizer can generalize the model to practical and real data. This opens up a horizon for applying the MKD algorithm on the subject of temporal downscaling. Other than this, the MKD algorithm can be a potent deconvolution algorithm to recover Gaussian blurring and generate high-resolution images keeping the mean pixel value of the image unchanged. There are still avenues to be explored using this algorithm. Further work is needed to show how the MKD algorithm's output affects the outcome of predictive machine learning models like Long Short Term Memory (LSTM), and prophet.

\section*{Acknowledgments}
The research by M. Kamrujjaman was partially supported by the University Grants Commission (UGC), the Ministry of Science and Technology,
and  the Bose Center for Advanced Study and Research in Natural Sciences, University of Dhaka.
\section*{Conflict of interest}
The authors declare no conflict of interest. 

\section*{Data sharing}
Data can be provided on a properly justified request.
\section*{Ethical approval}
No consent is required to publish this manuscript.

\section*{Author contributions}
Conceptualization,  MK and MAM; methodology, MAM and MK; software, MAM and MK; validation, MK; formal analysis, MAM; investigation, MK; resources, MK;  data curation, MAM; original draft preparation, MAM; review and editing, MK; supervision, MK. All authors have read and agreed to the published version of the manuscript.
%
   
    \nocite{*}
    \bibliographystyle{unsrt}
    \bibliography{reference}
    
     \newpage
    \appendix
    \raggedright \textbf{\Huge Appendix }
    
    \section{Additional Figures}\label{app:fig}
    \begin{center}
    	\includegraphics[width=\textwidth]{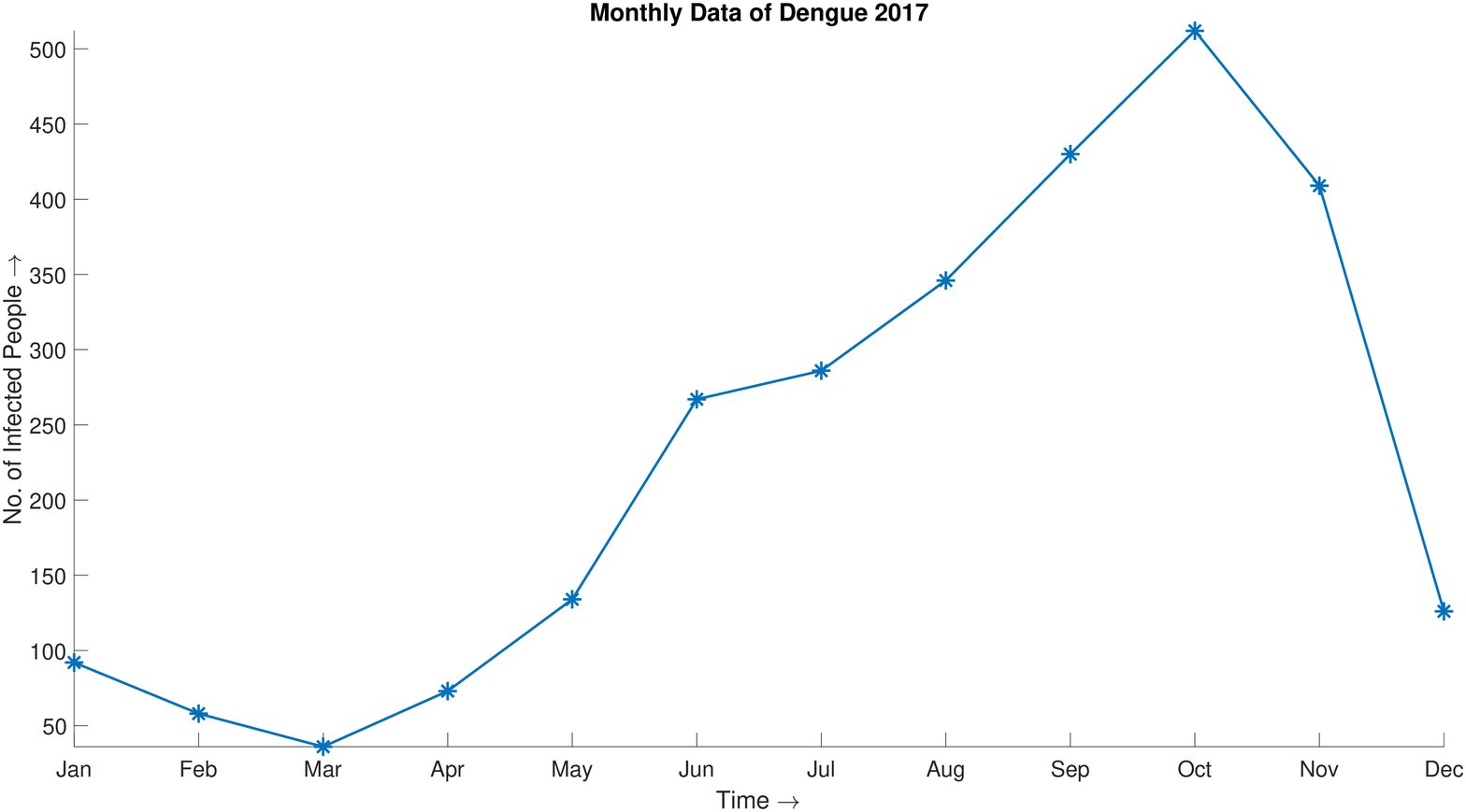}
    	\captionof{figure}{The monthly aggregate of the DENV infection in Bangladesh in the year 2017}
    	\label{fig:monhtly_distro}
    \end{center}
    
    \begin{center}
    	\includegraphics[width=\textwidth]{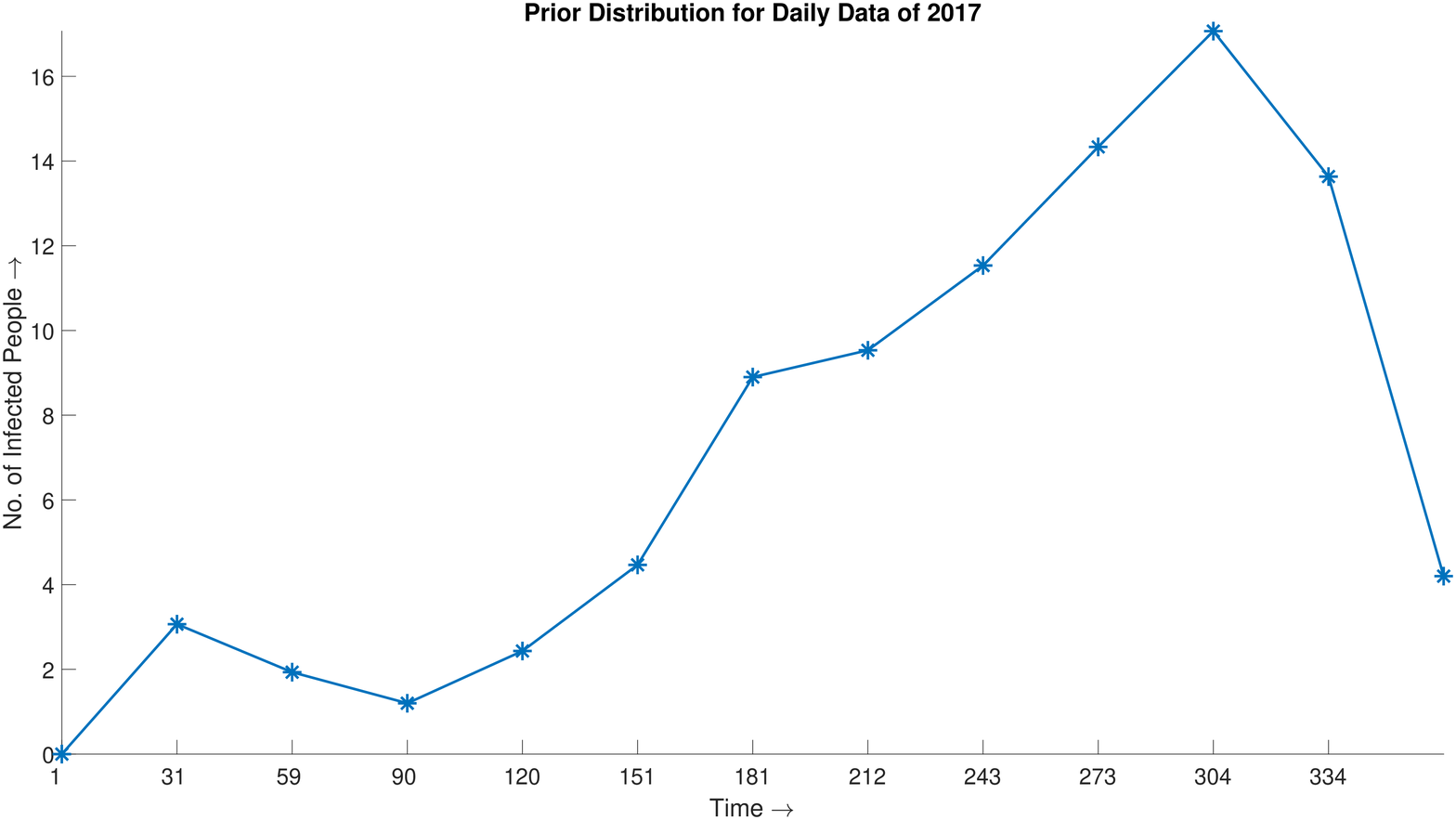}
    	\captionof{figure}{The prior distribution of the DENV infection of Bangladesh in the year 2017 generated from the monthly aggregate distribution exhibited in the figure \ref{fig:monhtly_distro}.}
    	\label{fig:prior_distro}
    \end{center}

    \begin{minipage}{0.5\textwidth}
    	
    	\centering
    	\includegraphics[width=\textwidth]{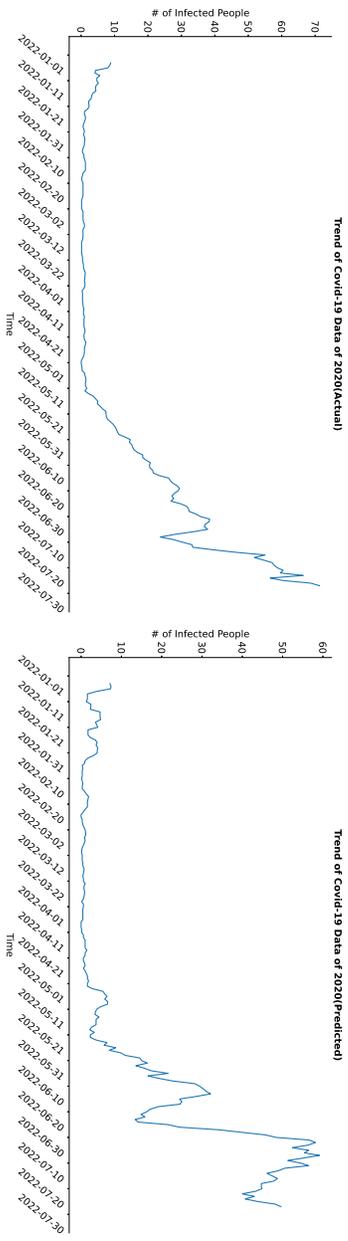}
    	\captionof{figure}{Trend of the actual dengue data.}
    	\label{fig:actualtrend}
    \end{minipage}%
    \begin{minipage}{0.5\textwidth}
    	\centering
    	
    	\includegraphics[width=\textwidth]{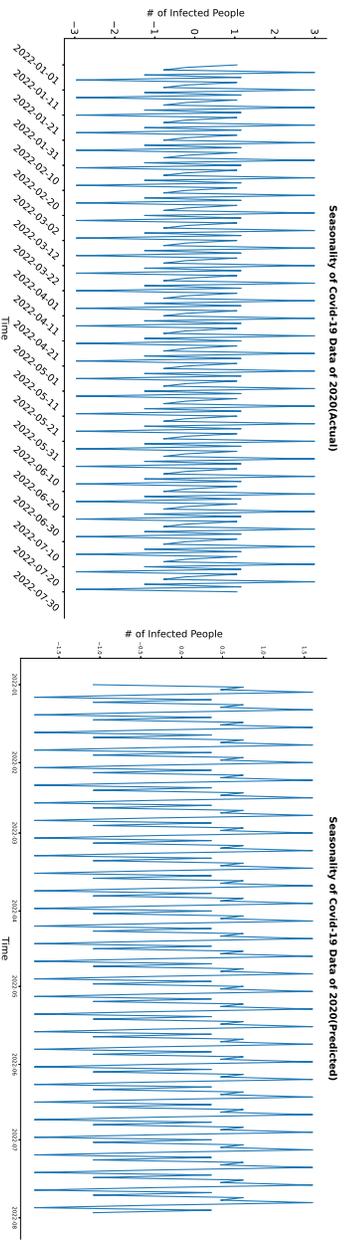}
    	\captionof{figure}{Trend of the synthetic dengue data.}
    	\label{fig:synthetictrend}

    \end{minipage}
    
    \vspace{1in}
    
    \begin{minipage}{0.5\textwidth}
    	
    	\centering
    	\includegraphics[width=\textwidth]{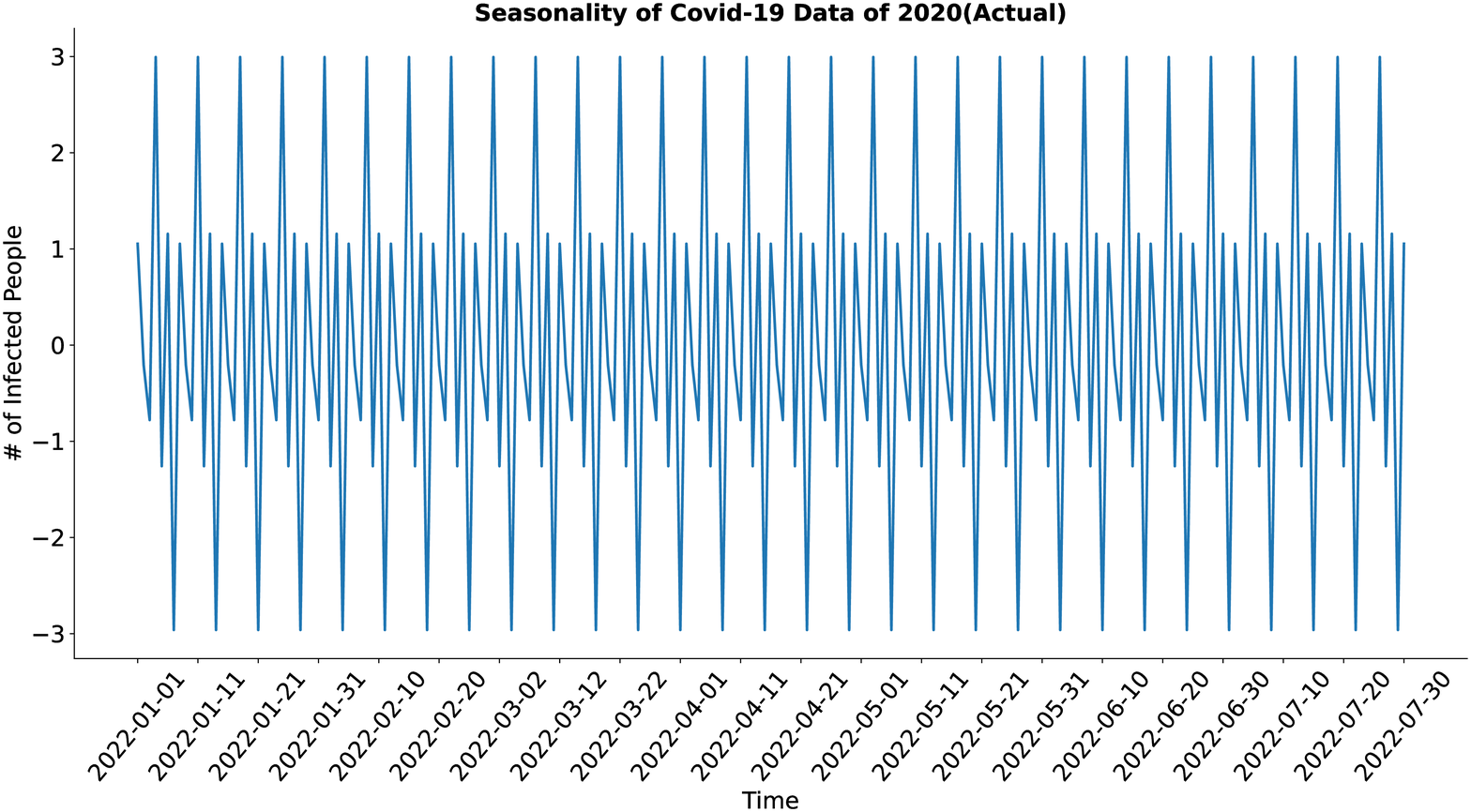}
    	\captionof{figure}{Seasonality of the actual dengue data.}
    	\label{fig:actualseason}
    \end{minipage}%
    \begin{minipage}{0.5\textwidth}
    	\centering
    	
    	\includegraphics[width=\textwidth]{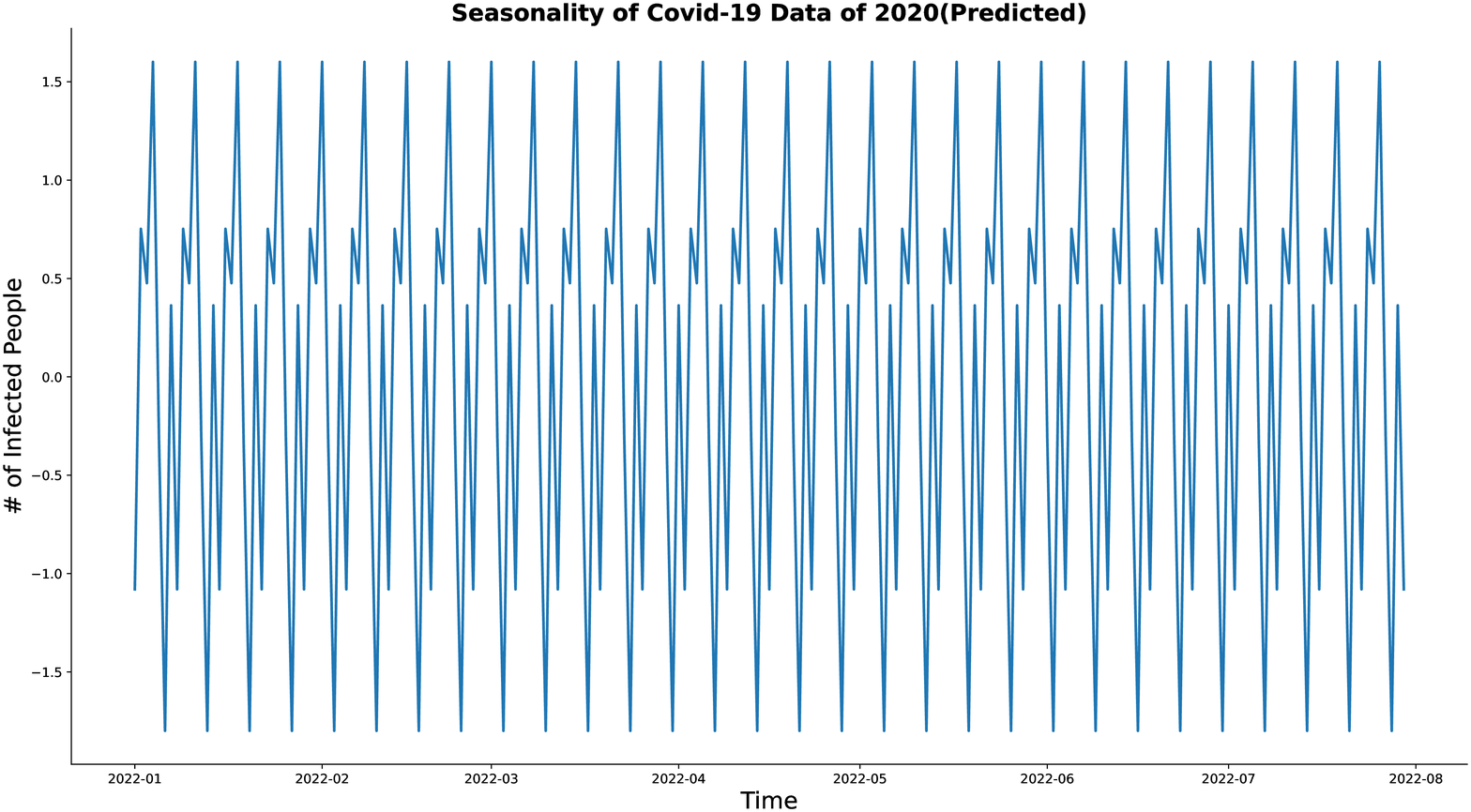}
    	\captionof{figure}{Seasonality of the synthetic dengue data.}
    	\label{fig:syntheticseason}

    \end{minipage}
    
    \vspace{1in}
    
    \begin{minipage}{0.5\textwidth}
    	
    	\centering
    	\includegraphics[width=\textwidth]{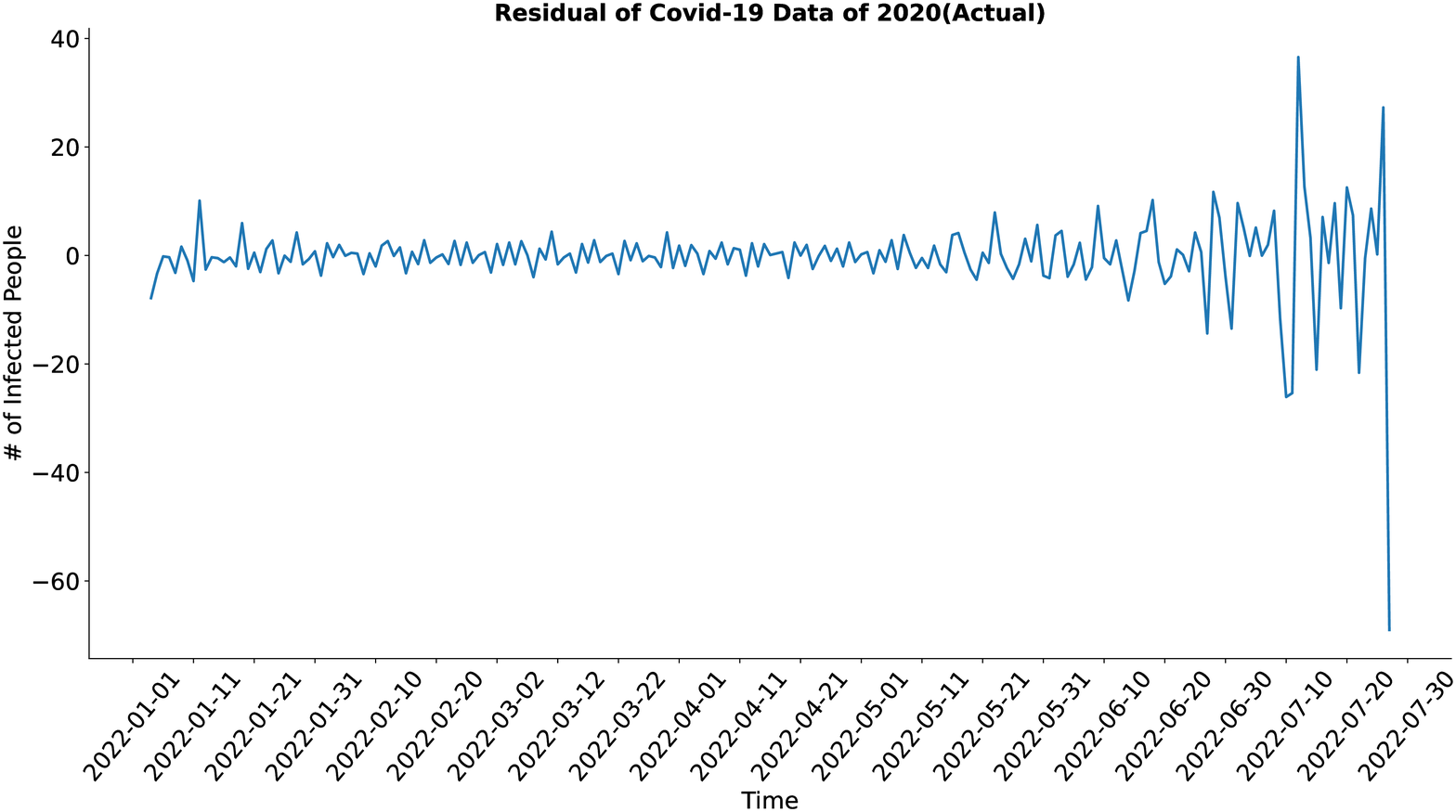}
    	\captionof{figure}{Residual of the actual dengue data.}
    	\label{fig:actualresidual}
    \end{minipage}%
    \begin{minipage}{0.5\textwidth}
    	\centering
    	
    	\includegraphics[width=\textwidth]{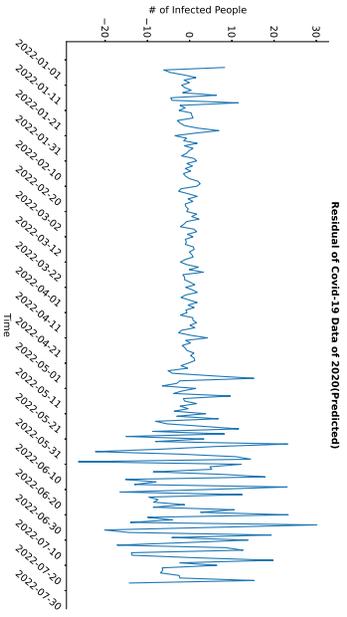}
    	\captionof{figure}{Residual of the synthetic dengue data.}
    	\label{fig:syntheresidual}

    \end{minipage}
    %
    
    %
    
    %

    \begin{minipage}{0.5\textwidth}
    	
    	\centering
    	\includegraphics[width=\textwidth]{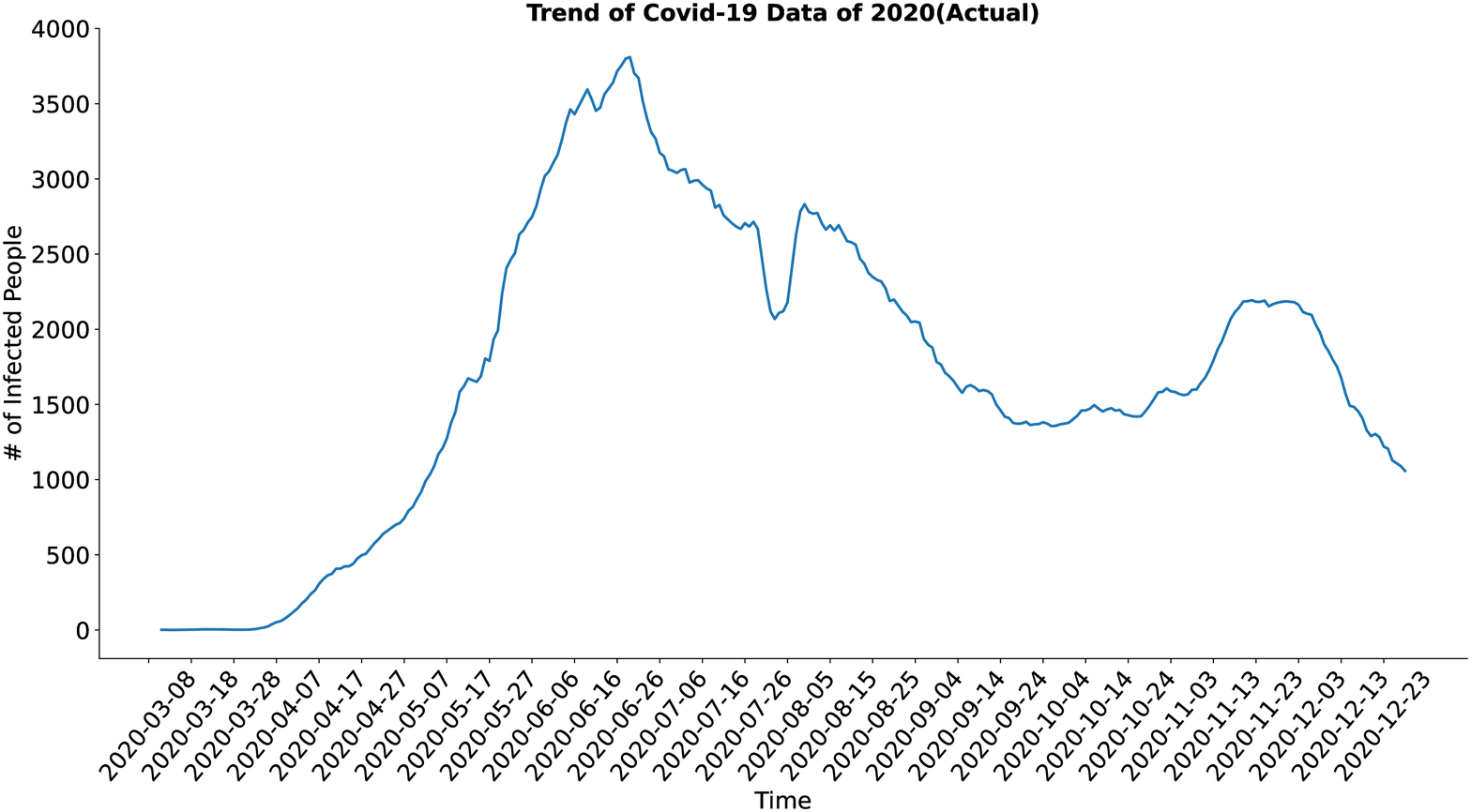}
    	\captionof{figure}{Trend of the actual COVID-19 data.}
    	\label{fig:actualtrend1}
    \end{minipage}%
    \begin{minipage}{0.5\textwidth}
    	\centering
    	
    	\includegraphics[width=\textwidth]{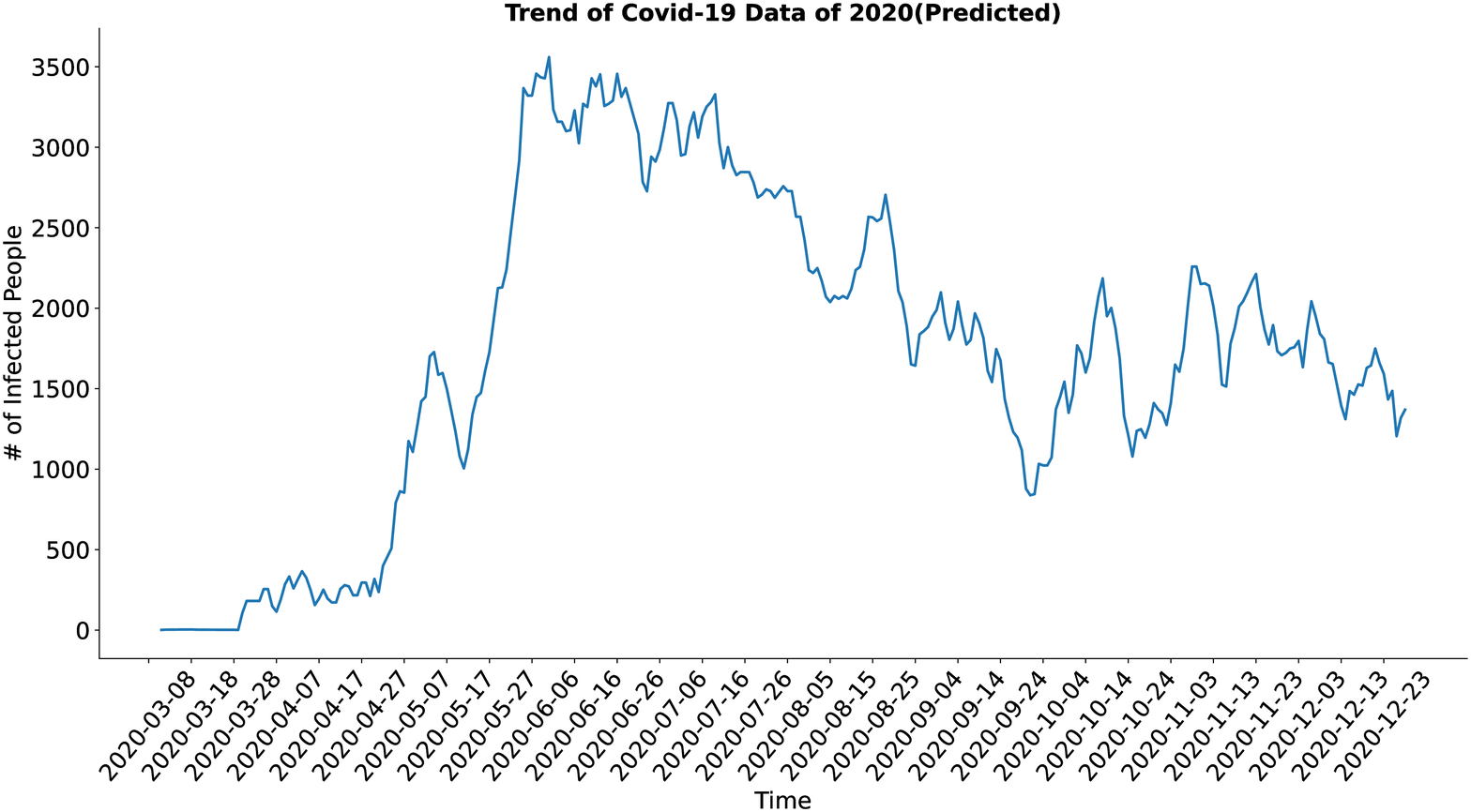}
    	\captionof{figure}{Trend of the synthetic COVID-19 data.}
    	\label{fig:synthetictrend1}

    \end{minipage}
    
    \vspace{1in}
    
    \begin{minipage}{0.5\textwidth}
    	
    	\centering
    	\includegraphics[width=\textwidth]{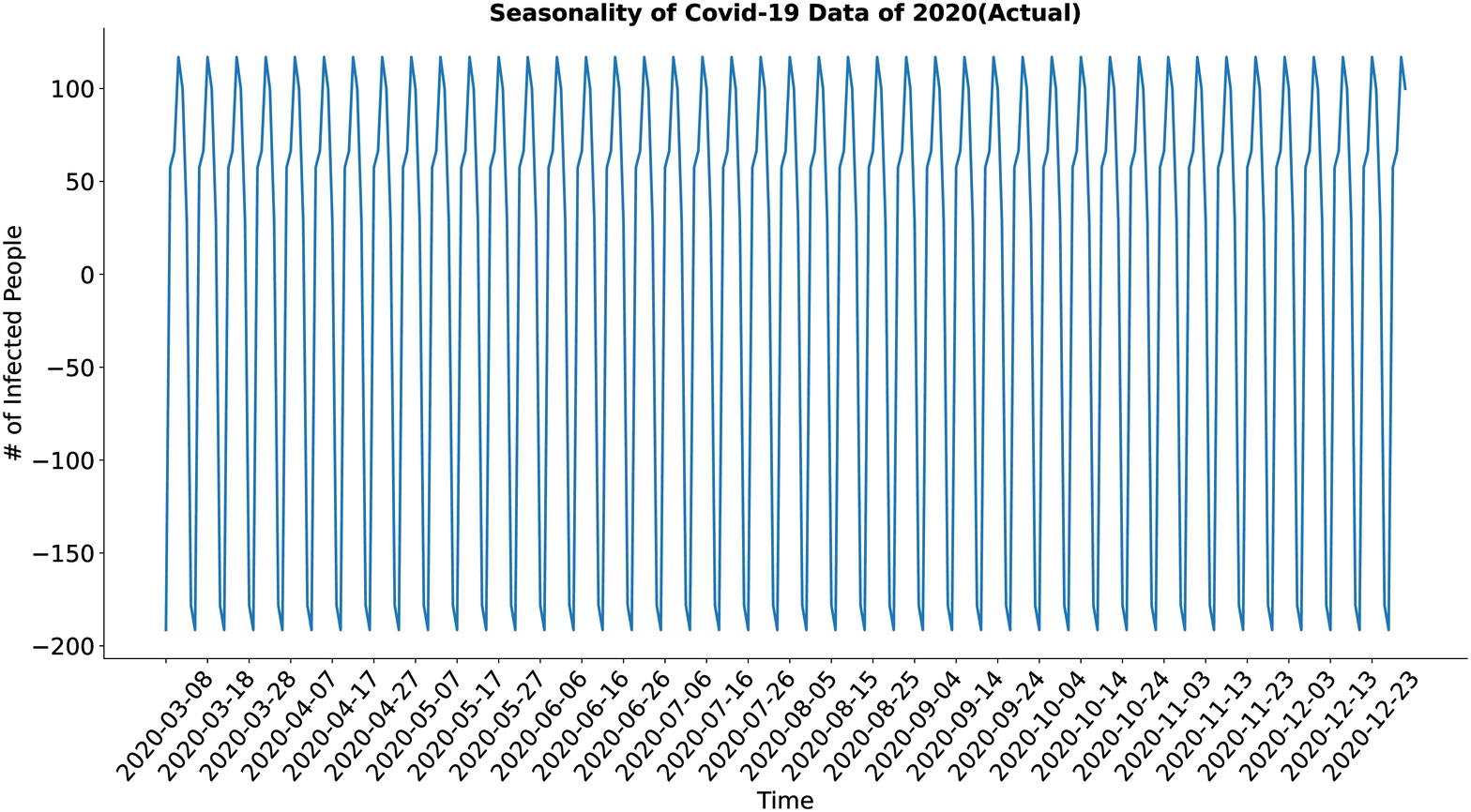}
    	\captionof{figure}{Seasonality of the actual COVID-19 data.}
    	\label{fig:actualseason1}
    \end{minipage}%
    \begin{minipage}{0.5\textwidth}
    	\centering
    	
    	\includegraphics[width=\textwidth]{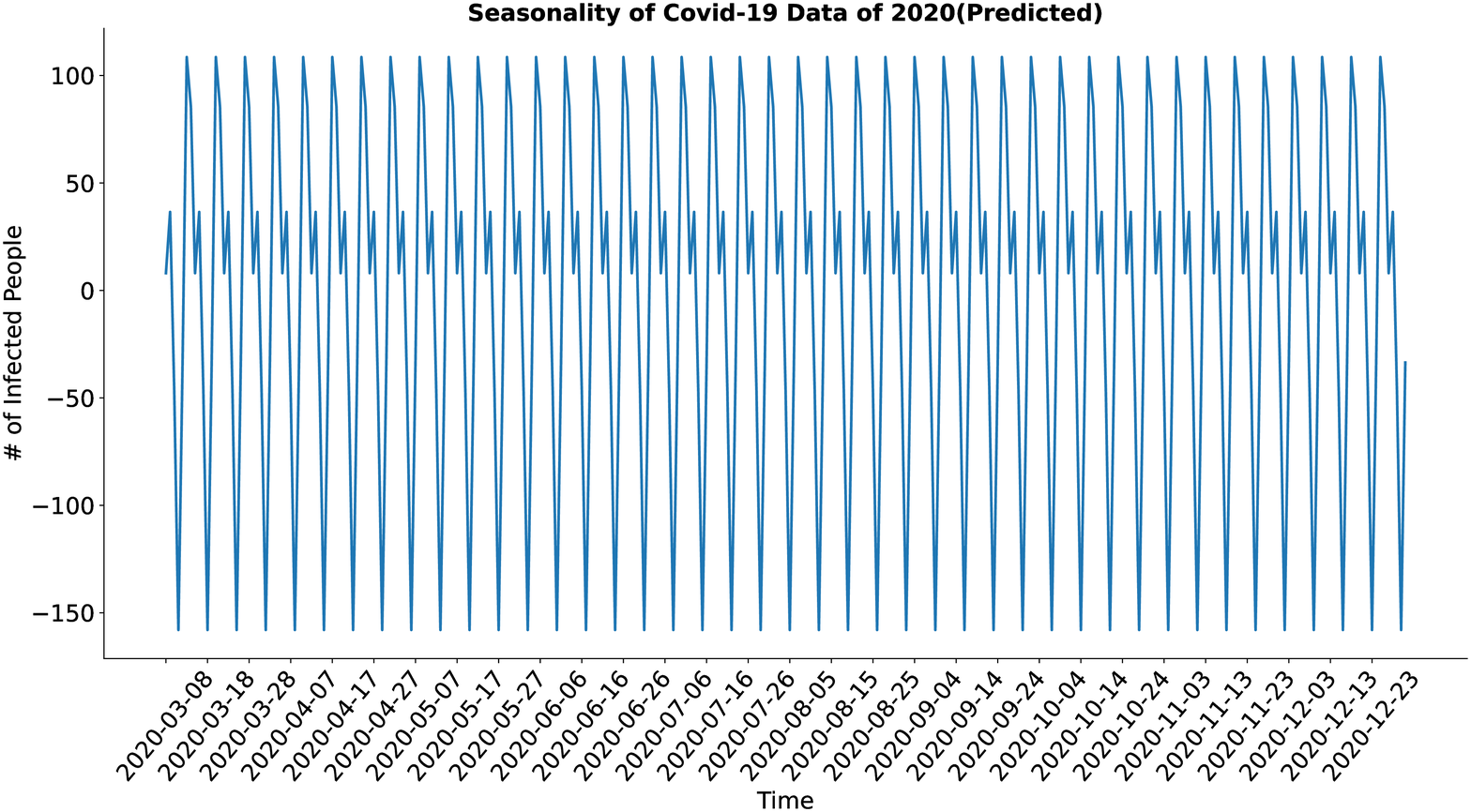}
    	\captionof{figure}{Seasonality of the synthetic COVID-19 data.}
    	\label{fig:syntheticseason1}

    \end{minipage}
    
    \vspace{1in}
    
    \begin{minipage}{0.5\textwidth}
    	
    	\centering
    	\includegraphics[width=\textwidth]{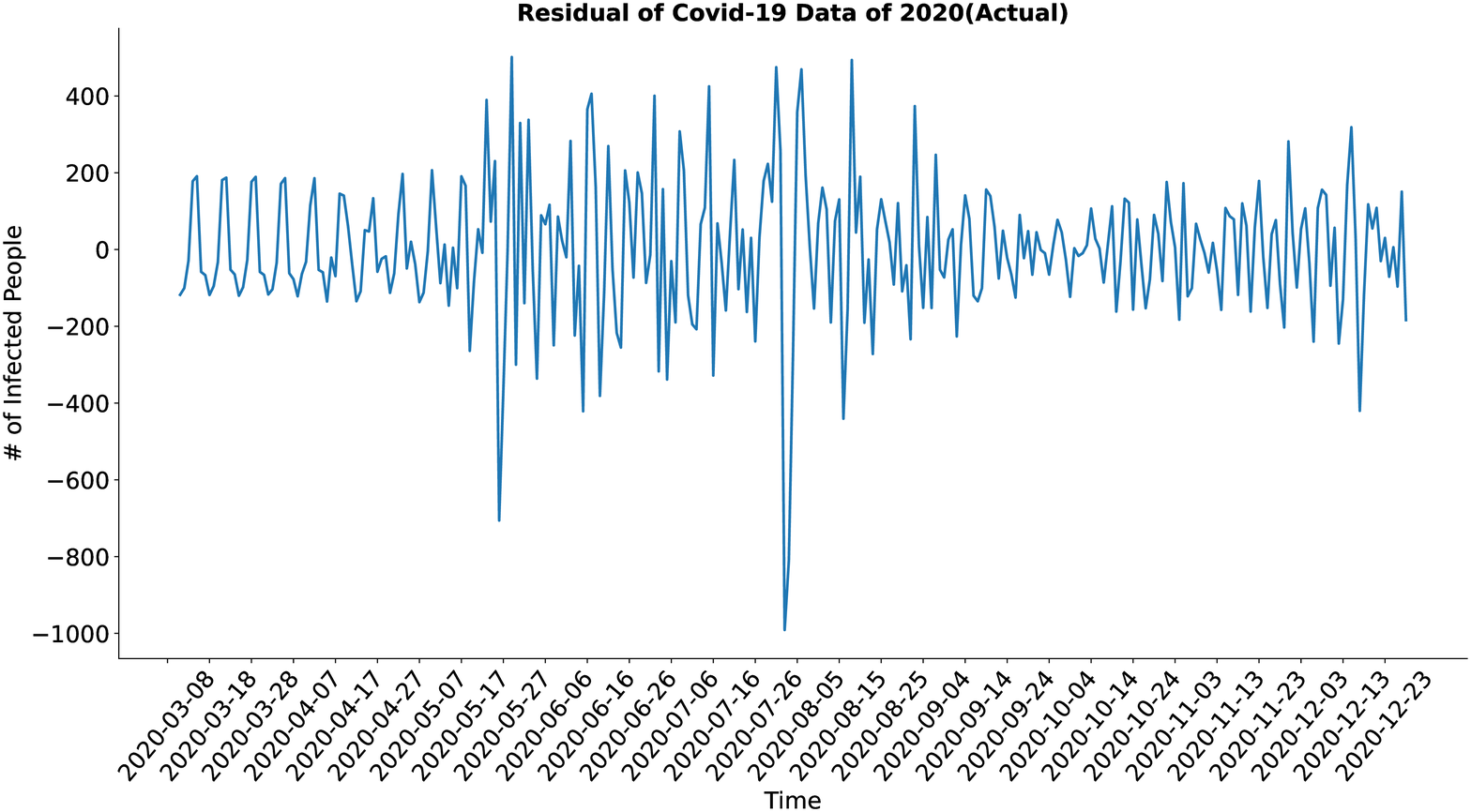}
    	\captionof{figure}{Residual of the actual dengue data.}
    	\label{fig:actualresidual1}
    \end{minipage}%
    \begin{minipage}{0.5\textwidth}
    	\centering
    	
    	\includegraphics[width=\textwidth]{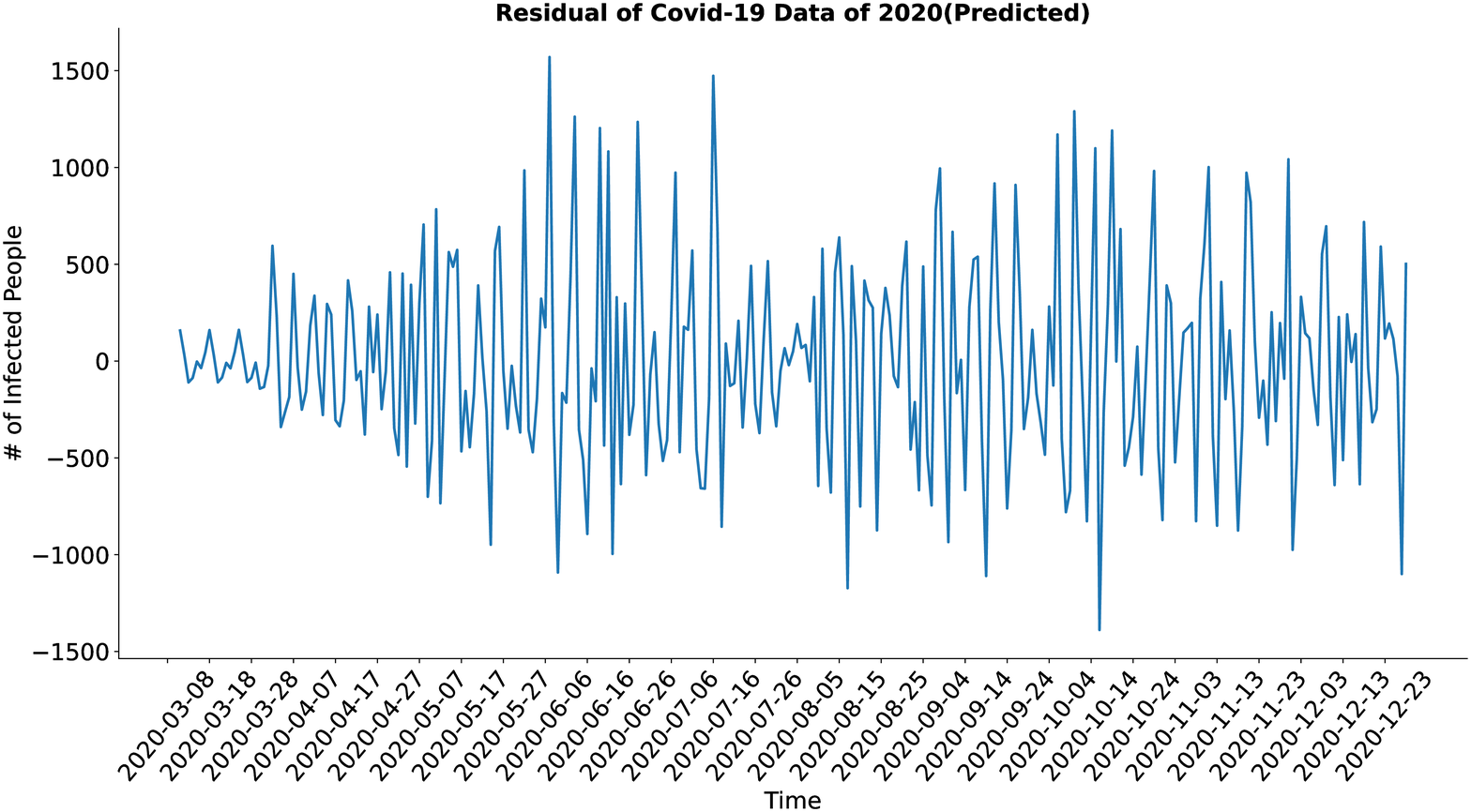}
    	\captionof{figure}{Residual of the synthetic COVID-19 data.}
    	\label{fig:syntheresidual1}

    \end{minipage}

    %
    %
    %
    %
    %
    
\end{document}